\documentclass[lettersize,journal]{IEEEtran}

\usepackage{cite}
\usepackage{amsmath,amssymb,amsfonts}
\usepackage{algorithm}
\usepackage{algpseudocode}
\usepackage{graphicx}
\usepackage{textcomp}
\usepackage{xcolor}
\usepackage{url}
\usepackage{multirow}
\usepackage{hhline}
\usepackage[caption=false,font=normalsize,labelfont=sf,textfont=sf]{subfig}
\usepackage{booktabs}
\usepackage{balance}
\usepackage{hyperref}

\def\BibTeX{{\rm B\kern-.05em{\sc i\kern-.025em b}\kern-.08em
    T\kern-.1667em\lower.7ex\hbox{E}\kern-.125emX}}
\begin{document}

\title{InterSim: Innovative Intersection Digital Twins for Human-in-the-Loop Driving Simulation}

\author {Luyang~Jiang,
        Chuheng~Wei,
        Yaguang~Li,
        Shaoqi~Wang,
        Yupeng~Zhou,
        Hyunsoo~Yun,
        Xishun~Liao,~\IEEEmembership{Member,~IEEE,}
        Brandon~Pitts,
        and~Ziran~Wang,~\IEEEmembership{Senior Member,~IEEE}


%
%
\thanks{This work was supported by the Indiana Department of Transportation (INDOT) Joint Transportation Research Program (JTRP) under Grant JTRP-4932.}%

\thanks{L. Jiang, C. Wei, Y. Li, S. Wang, Y. Zhou, H. Yun, B. Pitts, and Z. Wang are with the College of Engineering, Purdue University, West Lafayette, IN 47907, USA.

X. Liao is with the College of Engineering, the University of Central Florida, Orlando, FL 32816, USA.

Corresponding author: Z. Wang, e-mail: ziran@ieee.org.

This work has been submitted to the IEEE for possible publication. Copyright may be transferred without notice, after which this version may no longer be accessible.}}

\markboth{Under Review for IEEE Internet of Things Journal}%
{Jiang \MakeLowercase{\textit{et al.}}: InterSim: Innovative Intersection Digital Twins for Human-in-the-Loop Driving Simulation}
%



\maketitle


\begin{abstract}
Innovative intersection designs can improve roadway safety and operational efficiency, but their uncommon geometries may create challenges for driver comprehension during early deployment. Evaluating these designs through full-scale field implementation can also be costly. This paper presents \textit{InterSim}, a Unity game engine-based human-in-the-loop (HITL) driving simulator that constructs digital twins of site-specific innovative intersections in a virtual environment. \textit{InterSim} links the real world and the simulation world through two closed cycles: a digital twin cycle that builds each intersection twin from agency design and traffic data and returns evidence from simulation to the agency, and an HITL cycle in which human drivers or automated driving algorithms operate the ego vehicle under identical scenario conditions. The research capabilities of \textit{InterSim} are demonstrated through two case studies: a surrogate safety study compares post-encroachment time distributions between a Reduced Conflict Intersection and a standard intersection under different traffic volumes; an eco-driving study compares human drivers with a rule-based agent using vehicle-to-infrastructure information and reveals differences in emission and driving behavior. These studies demonstrate that \textit{InterSim} provides a reusable digital twin platform for investigating innovative intersection designs from diverse perspectives. Demo video and open source code can be accessed at \href{https://purduedigitaltwin.github.io/InterSim/}{https://purduedigitaltwin.github.io/InterSim/}.

\end{abstract}


\section{Introduction}
\label{sec:introduction}

\IEEEPARstart{A}{s} traffic demand continues to rise in urban road networks, standard four-leg signalized intersections increasingly serve as operational bottlenecks and primary locations of vehicular conflict \cite{Zlatkovic2019_PerformanceMatrices}. To address these challenges, transportation agencies have turned to innovative intersection designs, such as Displaced Left Turns (DLTs), Reduced Conflict Intersections (RCIs), and Diverging Diamond Interchanges (DDIs) \cite{innovative2022, hughes2010alternative}. By rerouting left turns or separating crossing maneuvers, these unconventional configurations reduce high-severity conflict points and simplify signal phasing \cite{MUT2022, steyn2014displaced}. Empirical field evaluations demonstrate that such geometric modifications can enhance operational throughput and reduce angle collisions compared to traditional layouts \cite{inman2012field, hummer2014restricted}.

Realizing the full benefits of innovative intersections requires rigorous evaluation across both microscopic and macroscopic dimensions. Unfamiliar travel paths and non-standard visual guidance can induce driver hesitation and confusion during early deployment \cite{Hughes2006_InnovativeIntersectionSafety, wei2024dilemma,shumaker2013barriers}. Evaluating how human drivers perceive new road geometries and lane markings is critical to prevent human error from undermining intersection safety. At the macroscopic level, infrastructure owner-operators (IOOs) must assess how these intersections perform across different traffic volumes and surrogate safety measures, as well as their impact on the existing road network \cite{abdelrahman2020evaluation, alkaissi2022traffic}. Traditionally, agencies often rely on pilot installations or field observations to evaluate these outcomes. However, such trial-and-error approaches in the real world are costly and expose road users to unnecessary safety risks \cite{gartner2025scalable, zhong2019alternative}.

To evaluate candidate designs without physical disruption, virtual simulation offers a safe and controlled alternative. However, existing tools still fall short in a different way: microscopic traffic simulators, such as SUMO \cite{sumo2018}, effectively evaluate operational capacity and delay, but cannot capture human driver behavior; conversely, autonomous driving simulators such as CARLA \cite{carla2017} deliver high visual fidelity, yet prioritize autonomous agent algorithms over human-in-the-loop (HITL) interaction. To bridge those gaps, some frameworks utilize digital twins to connect the physical world with virtual models \cite{wu2025digital, irfan2024toward}, but current implementations mainly focus on post-construction asset management, health monitoring, or visualization alone \cite{ge2024digital, wu2025digital}. They rarely provide a workflow to evaluate pre-construction blueprints through interactive driver feedback. Consequently, IOOs like transportation agencies still lack an integrated and realistic platform to evaluate intersection designs through human driving interactions.

To bridge this gap, we introduce \textit{InterSim}, a Unity game engine-based driving simulator that constructs digital twins of innovative intersections for HITL evaluation. \textit{InterSim} connects the physical and virtual environments through two closed cycles: a digital twin cycle and an HITL cycle. In the digital twin cycle, road and traffic data provided by IOOs are used to build virtual replicas, and the analysis results can then be returned to IOOs to guide design adjustments. In the HITL cycle, human drivers or automated driving algorithms operate the ego vehicle within the twin in real time, interacting with background traffic and signals. To support an extensive range of experiments, the platform also supports full customization of vehicle dynamics, sensors, and vehicle-to-everything (V2X) communication.

We demonstrate the practical utility of \textit{InterSim} through two HITL case studies: a surrogate safety evaluation between an RCI and a standard intersection across multiple traffic demands, and an eco-driving algorithm evaluation comparing human driver emissions against an eco-driving agent using vehicle-to-infrastructure (V2I) communication at a DLT intersection. The primary contributions of this paper are summarized as follows:
\begin{itemize}
    \item
    We develop \textit{InterSim}, a Unity game engine-based driving simulator with open-source release. It introduces a novel digital twin architecture that reconstructs innovative intersections from the design and traffic data of IOOs and returns the analysis to them, so that a design can be evaluated and refined before construction.

    \item
     \textit{InterSim} provides a configurable HITL framework that integrates vehicle dynamics, traffic generation, sensor emulation, V2X communication, and physical devices to perceive realistic driver behavior.
    
    \item
    We conduct two HITL case studies: a surrogate safety comparison between an RCI and a standard intersection, and an emission comparison between human drivers and an eco-driving agent informed by V2I. In both studies, the HITL cycle records how drivers perceive and respond to the intersection, and the digital twin cycle turns these records into evidence that agencies can use to evaluate and improve its design.
\end{itemize}

\section{Related Work}
\label{sec:related work}
Driving simulation has evolved into a critical methodology for evaluating both vehicle systems and human driver behaviors in a safe, controlled environment. Existing simulation platforms generally fall into distinct categories based on their primary applications. However, evaluating human driver comprehension and compliance at complex, innovative intersections requires a unique combination of high-fidelity environmental modeling and low-latency HITL capabilities. This section reviews existing simulation paradigms and relevant work evaluating innovative intersections.

\subsection{Simulators and Digital Twins}
Microscopic traffic simulators~\cite{sumo2018, vissim, Aimsun} model individual vehicle trajectories with behavioral algorithms to evaluate aggregate traffic operations~\cite{Olstam2004ComparisonOC, ahmed2021review, barcelo2005microscopic}. Recent work couples SUMO with language model agents that tune scenarios against network metrics~\cite{li2026chatsumo} and derives surrogate safety conflicts from VISSIM trajectories in connected and automated vehicle (CAV) scenarios~\cite{sauciur2025evaluating}. However, vehicles in these platforms are modeled strictly as algorithmic entities rather than human-controlled agents. Consequently, they cannot capture how human drivers perceive and navigate intersections, and their abstract road networks cannot demonstrate visually immersive scenarios that drivers see in the real world.

Autonomous driving simulators, such as CARLA~\cite{carla2017}, LGSVL~\cite{rong2020lgsvl}, and AirSim~\cite{shah2017airsim, wei2025integrating}, leverage commercial game engines to render urban scenes and emulate perception sensors, including cameras, LiDAR, and radar. These platforms serve primarily as closed-loop testbeds for benchmarking learning-based autonomous driving policies~\cite{yang2026raw2drive, ostreika2025deep}. Although human control inputs can be introduced, integrating physical control devices and configuring data pipelines require substantial effort. Moreover, incorporating unconventional intersections in cumbersome simulators often requires extra work.

To achieve greater geometric and interactive flexibility, recent studies utilize commercial game engines directly to build customized simulation environments for CAV research and digital twin prototyping~\cite{unityML2023, wang2021}, demonstrating that game engine foundations naturally accommodate complex 3D roadway geometries, customizable physics, and interactive user controls. In parallel, digital twins in transportation pair a physical system with a virtual model synchronized by field data~\cite{wang2022mobility, irfan2024toward}. Wang et al.~\cite{wang2021digital} build a digital twin to guide human drivers through non-signalized intersections alongside connected vehicles, while the surveys by Ge and Qin~\cite{ge2024digital} and Wu et al.~\cite{wu2025digital} report digital twins used to monitor and maintain roads after construction. These digital twins are updated with data from roads and vehicles already in service for post-construction analysis. However, limited work has focused on digital twins that are built before field deployment, which allows traffic agencies to test road designs in advance to mitigate safety risks.

\begin{figure*}[t]
    \centering
    \includegraphics[width=1\linewidth]{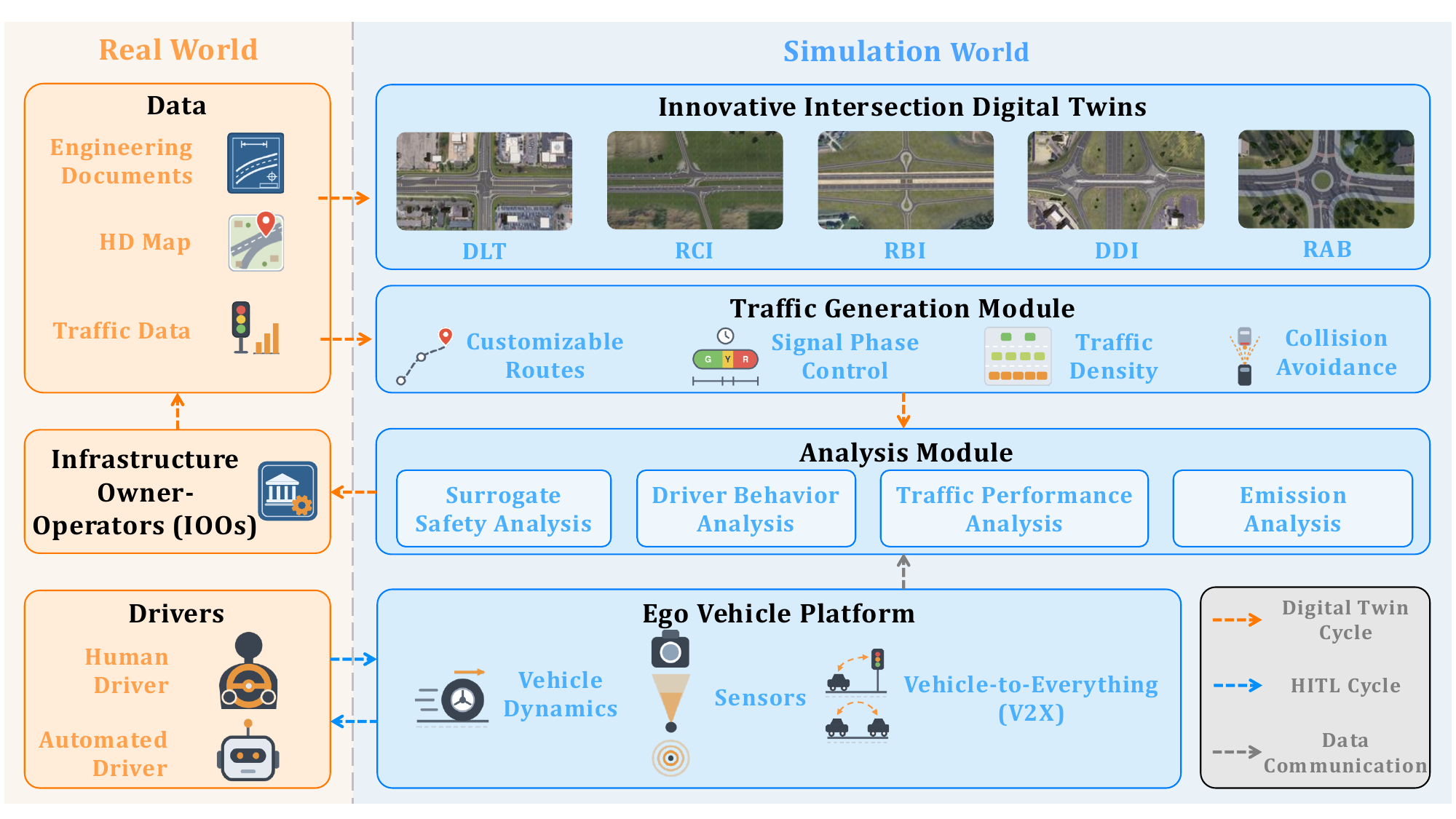}
    \caption{Overall architecture of the driving simulator for innovative intersection designs: Displaced Left Turn (DLT), Reduced Conflict Intersection (RCI), Roundabout Interchange (RBI), Diverging Diamond Interchange (DDI), Two-Lane Roundabout (RAB).}
    \label{fig:architecture}
\end{figure*}

\subsection{Evaluation of Innovative Intersection Designs}
By reconfiguring turning movements and signal phasing, innovative intersections such as DLTs, RCIs, and DDIs alter how traffic flows through a junction~\cite{hughes2010alternative,steyn2014displaced,hummer2014restricted}. 
Existing evaluations primarily assess whether a candidate layout is geometrically feasible and how it affects macroscopic traffic performance and conflict frequency. 
The Federal Highway Administration (FHWA) provides comparative frameworks for benchmarking these alternative geometries against conventional intersections~\cite{hughes2010alternative}, in which DLT and RCI guides only address practical implementation details ranging from lane markings to human factors~\cite{steyn2014displaced,hummer2014restricted}.

Field and simulation studies further examine the performance of innovative intersections. Abdelrahman et al. \cite{abdelrahman2020evaluation} evaluate DLT intersections using before-after, cross-sectional, and high-resolution traffic-data analyses, while Qu et al. \cite{qu2020statistical} and Inman and Haas \cite{inman2012field} study DLT and RCI safety using crash records and field observations. When field deployment is costly, simulation-based studies commonly use tools such as VISSIM to evaluate operational and safety performance, as shown in DLT, unconventional U-turn, and RCI studies \cite{alkaissi2022traffic,xiang2016evaluating,molan2022framework}. Industrial safety tools, such as the Surrogate Safety Assessment Model (SSAM) \cite{gettman2008surrogate}, further support safety inference from simulated trajectories.

While these field and simulation studies capture aggregate traffic and safety outcomes, driver comprehension remains less systematically addressed. FHWA studies examine DLT marking and DDI navigation using driver-centered evaluations \cite{inman2009sign,inman2007drivers}. In addition, Sun et al. \cite{sun2017driving} use a driving simulator to study J-turn lanes and U-turn spacing, while Shumaker et al. \cite{shumaker2013barriers} identify unfamiliarity and public acceptance as major deployment barriers. 
Overall, prior work tends to evaluate traffic operations and driver behavior separately, leaving a need for a reusable platform that places human drivers directly within site-specific intersection environments.

\section{System Architecture}
\label{sec:architecture}
This section introduces the architecture that couples the real world with the simulation world for the study of innovative intersection designs, as shown in Fig.~\ref{fig:architecture}. In the real world, IOOs hold the engineering documents, maps, and traffic data that describe each facility, and drivers are the road users who operate vehicles at the site. In the simulation world, \textit{InterSim} maintains a digital twin of each intersection, generates the traffic that populates it, provides the ego vehicle that the driver operates, and analyzes the data produced by each run. Two closed cycles connect the two worlds. The digital twin cycle transfers design and traffic information from the IOOs into the twin and returns analysis results to them, while the HITL cycle connects the driver to the ego vehicle platform in real time. A third flow logs the states of the ego vehicle for analysis and thereby links the two cycles. This organization separates the information that defines a scenario from the interaction that takes place inside it.

\subsection{Digital Twin Cycle}
The digital twin cycle begins with the data that IOOs, such as state and local transportation agencies, maintain for each facility. Engineering documents provide lane configurations, pavement markings, traffic signs, and signal positions. High-definition (HD) maps extracted from aerial imagery provide road geometry together with the buildings, infrastructure, and vegetation that surround the site. Traffic data provide traffic volumes and signal timing plans. These sources are complementary: the documents define how the intersection is designed to operate, the maps define where its elements are located, and the traffic data define the demand that it serves.

From these inputs, we construct a digital twin of each intersection, that is, a virtual replica whose geometries and surroundings correspond to the intersection in the real world. Each twin preserves road geometry, pavement markings, traffic signs and signals so that a driver encounters the same visual guidance and maneuver sequence as the real site provides. \textit{InterSim} currently provides five innovative intersection twins: a DLT, an RCI, a DDI, a Roundabout Interchange (RBI), and a Two-Lane Roundabout (RAB). The same procedure applies to the standard intersection used as a comparison baseline in Section~\ref{sec:case study}, and the construction steps are described in Section~\ref{sec:workflow}.

Traffic data enter the twin through the traffic generation module, which populates the scenario with surrounding vehicles. Customizable routes define the waypoint paths that vehicles follow through the intersections, and traffic signals are set up according to the timing plans of the site. Traffic density is also fully customizable for each route, which is explained in more detail in Section~\ref{sec:workflow}, while real-world traffic data can also be imported to simulate more realistic traffic flow. Collision avoidance governs car-following, stopping, and blind-spot responses so that the interaction between the vehicles and the ego vehicle follows natural behavior. With all these customizable settings and data, the drivers are able to encounter traffic flows and conditions of the real intersection rather than a generic scenario.

The cycle closes through the analysis module. Each run logs vehicle trajectories, driver inputs, signal states, and sensor readings. Currently, \textit{InterSim} supports four major types of analysis: surrogate safety analysis reports conflict-based measures such as post-encroachment time (PET); driver behavior analysis characterizes speed and acceleration; traffic performance analysis reports travel time, delay, and throughput; emission analysis estimates energy consumption and pollutant emissions. The results are returned to the IOOs with deeper analysis and evaluation, who can revise a candidate design in many different ways, such as road geometry, signing, marking, or signal timing; the revised versions can be reapplied in the digital twin and go through another round of evaluation. This cycle allows design alternatives to be examined before actual deployment, which is more costly and difficult in the real world.

\subsection{Human-in-the-Loop Cycle}
The HITL cycle connects the drivers in the real world with the ego vehicle platform in the simulation world, and two types of drivers share the same connection point. A human driver operates the ego vehicle through a physical steering wheel and pedal set, or through a keyboard or gamepad when a physical wheel is not practical, and receives visual and steering feedback from the running scenario. Alternatively, one can also plug in a rule-based driving agent or learning-based autonomous driving algorithm that reads the same vehicle states and issues the same throttle, brake, and steering commands, as illustrated by the eco-driving agent in Section~\ref{sec:case study}. Because both driver types attach to the architecture at the same point, the same twin, traffic settings, and logging pipeline can be used to compare human with automated driving under identical scenario conditions.

The ego vehicle platform with various modules provides high flexibility and a broad range of experiments. Vehicle dynamics reproduce the motion of the ego vehicle through a physics model that can be tuned to a specific production vehicle. Sensors emulate the cameras, LiDAR, and radar that provide perception inputs to automated drivers and sensor records to the analysis module. The V2X module exchanges messages with roadside units and neighboring vehicles~\cite{wei2026cooperative}, such as the signal phase and timing (SPaT) information used in Section~\ref{sec:case study}. Table~\ref{tab:customizable_features} summarizes the configurable elements of \textit{InterSim}.

\begin{table}[!t]
    \centering
    \caption{Core Customization Capabilities of \textit{InterSim}}
    \label{tab:customizable_features}
    \footnotesize
    \setlength{\tabcolsep}{4pt}
    \renewcommand{\arraystretch}{1.15}
    \begin{tabular}{
        p{0.2\linewidth}
        p{0.7\linewidth}
    }
        \hline
        \textbf{Category} &
        \textbf{Representative Configurable Elements} \\
        \hline

        Map and Scene &
        Road models (five built in), traffic signs, signals, vegetation, buildings, and any 3D models \\

        Traffic &
        Routes, traffic volume, signal timing, and real-world traffic data import \\

        Vehicles &
        Vehicle models (four built in), dynamics, surrounding vehicle behavior, and algorithms \\

        Sensors &
        Radar, LiDAR, and cameras with configurable placement, field of view, 
        range, resolution, and update rate \\

        \hline
    \end{tabular}
\end{table}

\section{Real-to-Sim Workflow}
\label{sec:workflow}
\begin{figure}[!t]
    \centering
    \includegraphics[width=0.7\linewidth]{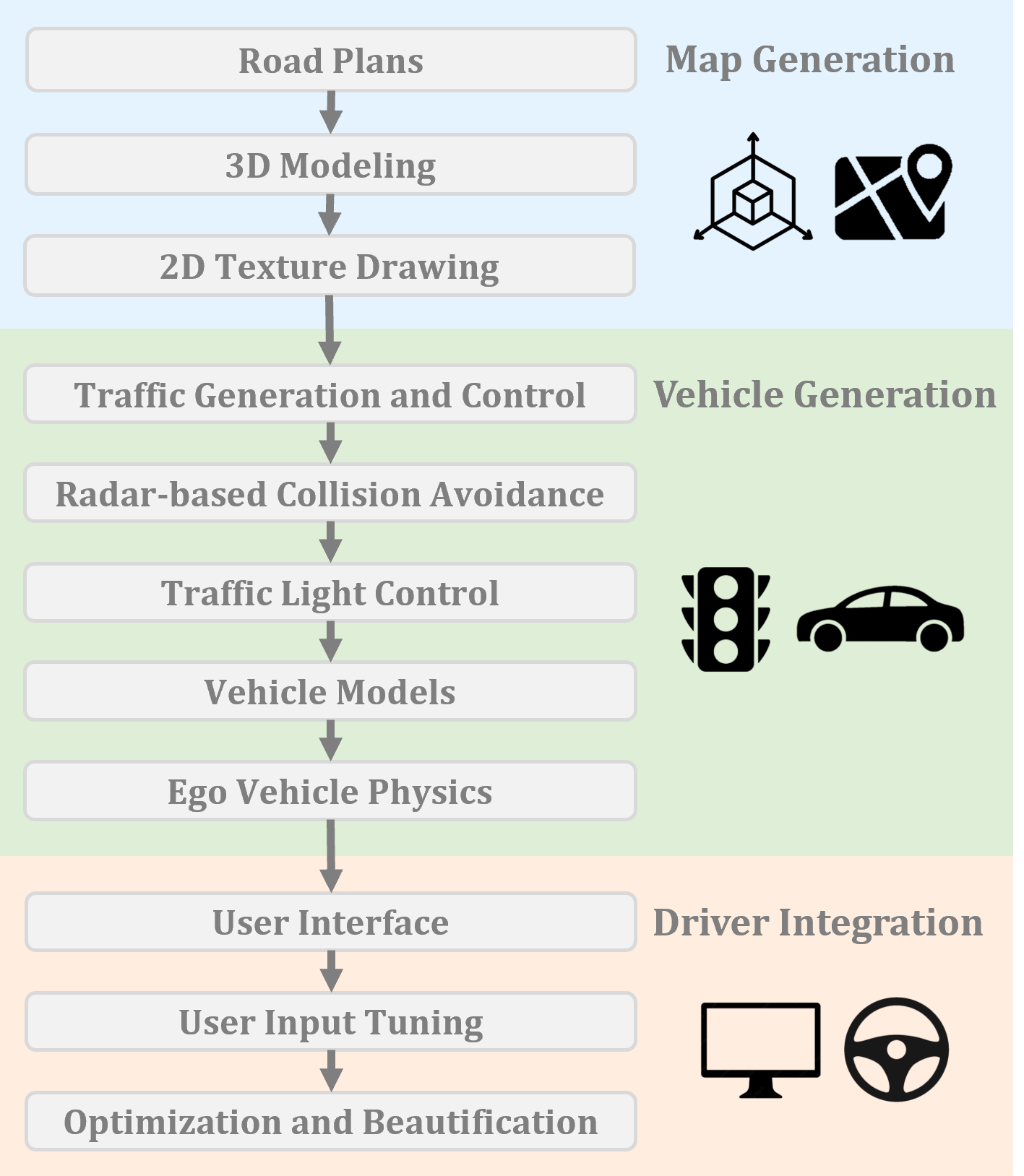}
    \caption{Real-to-sim workflow of the driving simulator.}
    \label{fig:workflow}
\end{figure}

This section explains the workflow for building each scenario from real intersection layouts to functional components in the simulation world. As shown in Fig.~\ref{fig:workflow}, the workflow includes three parts: 1) The first part generates the map and roadway geometry from real drawings and aerial imagery; 2) The second part generates vehicle traffic through waypoint paths, traffic logic, and motion settings; 3) The third part configures driver interaction for the ego vehicle through keyboard, gamepad, or steering-wheel input. Finally, metrics that evaluate the performance of the simulator are summarized.

\subsection{Map Generation}
\begin{figure}[!t]
        \centering
        \subfloat[]{%
            \includegraphics[width=0.8\linewidth]{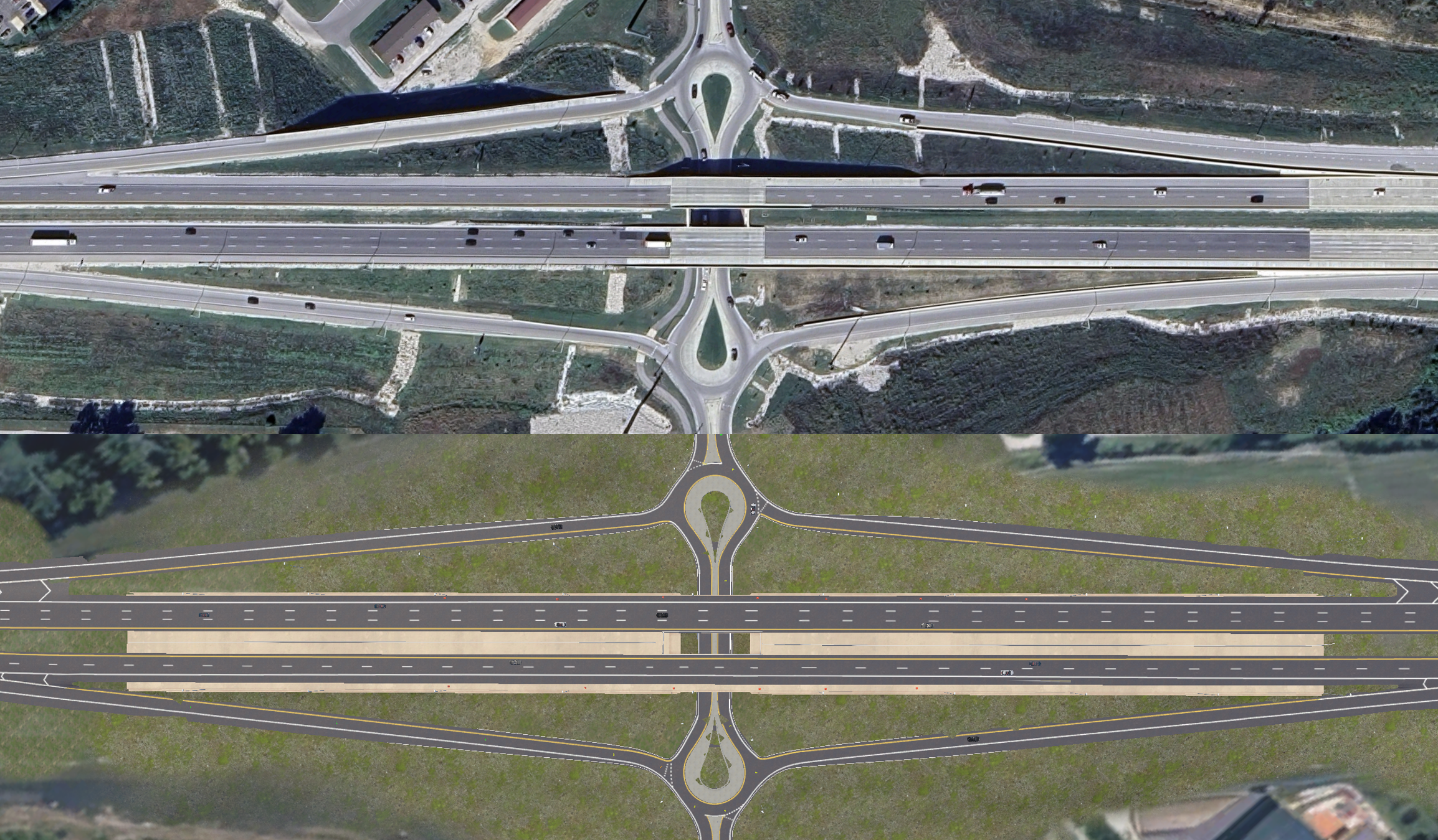}%
            \label{fig:map-rbi}}
        \\[0.5em]
        \subfloat[]{%
            \includegraphics[width=0.8\linewidth]{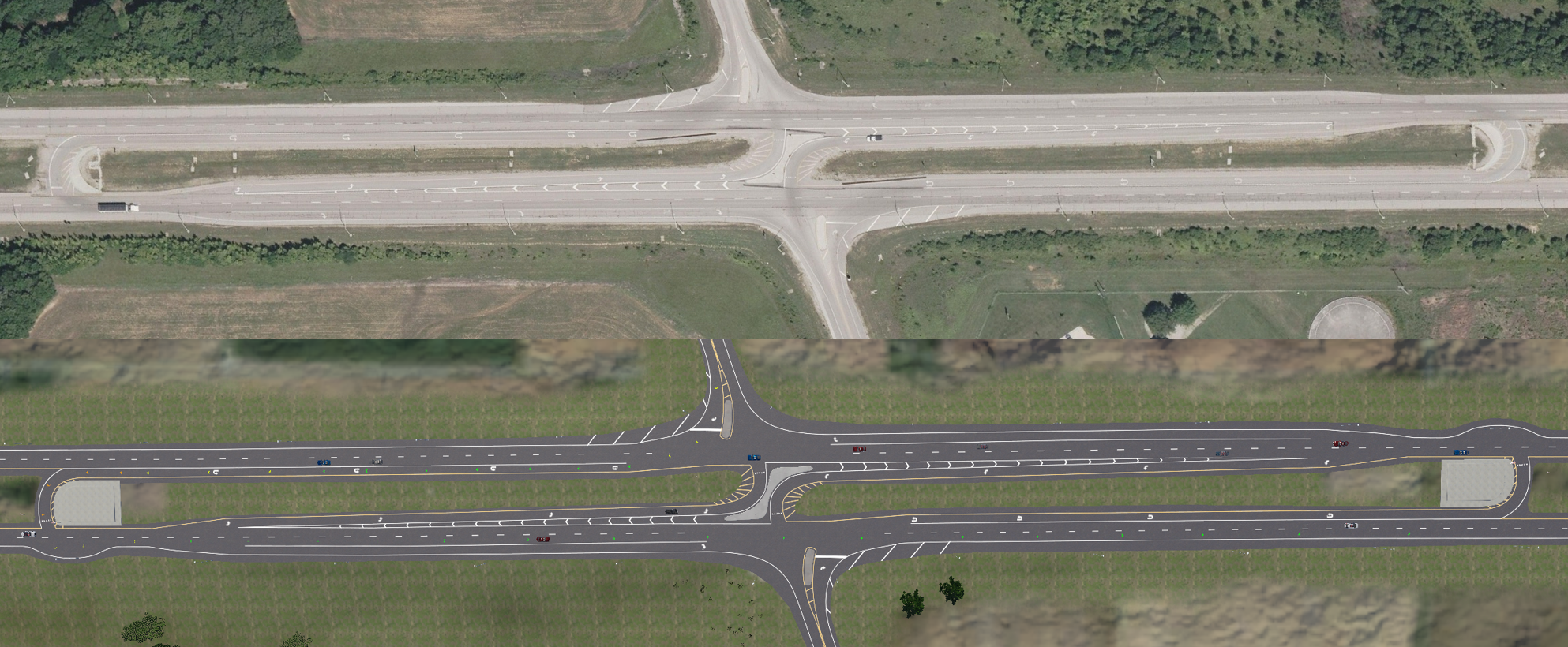}%
            \label{fig:map-rci}}
        \caption{Maps in the real world and simulation world: (a) RBI and (b) RCI.}
        \label{fig:map-comparison}
    \end{figure}

    Each scenario begins with the construction of the map that reflects the geometry of the real intersection. We obtain engineering road-plan documents from state transportation agencies and use them as the basis for reproducing the roadway layout. We use these plans to guide the creation of 3D models in Blender, a 3D modeling software commonly used for geometric reconstruction and scene development \cite{blender}. In Blender, we model the roads and related structures according to the measurements in the documents, preserving the scale and alignment of the real site. We then create the environment's visual appearance by generating textures in Photoshop and applying them to the models \cite{photoshop}. After preparing these assets, we import them into Unity and assemble them into a complete simulation map. Two example maps are shown in Fig.~\ref{fig:map-comparison}. This workflow establishes a direct transition from the real-world layout to the simulation environment.

\subsection{Vehicle Generation}    
   Vehicle generation covers the creation of surrounding traffic, the behavior of surrounding vehicles, the logic for signalized movements, and the configuration of the ego vehicle used by the driver. The process begins with the definition of traffic flows that follow real-world patterns. Users specify a base flow rate and adjust it with multipliers for density, time of day, day of week, and path preferences. The final adjusted traffic flow rate (veh/h) $Q_{\mathrm{adj}}$ is computed as: 
    \begin{equation}
        Q_{\mathrm{adj}} =
        \delta_{\mathrm{g}}
        \,\delta_{\mathrm{tod}}
        \,\delta_{\mathrm{dow}}
        \,\delta_{\mathrm{p}}
        \, q_{0}
        \, L
        \, \eta_{\mathrm{cap}}
        \left( \frac{v_{\mathrm{p}}}{30} \right)^{\alpha},
        \label{eq:adjusted-flow}
    \end{equation}
    where $\delta_{\mathrm{g}}$, $\delta_{\mathrm{tod}}$, $\delta_{\mathrm{dow}}$, and $\delta_{\mathrm{l}}$ are the global density, time-of-day, day-of-week, and lane density multipliers. The term $q_{0}$ is the base flow rate in vehicles per hour per lane (veh/h/l), $L$ is the lane count, and $\eta_{\mathrm{cap}}$ denotes the lane capacity factor accounting for geometric and operational constraints. The variable $v_{\mathrm{p}}$ is the speed of vehicles defined by the user in m/s, and $\alpha$ is the speed-flow exponent used to model the nonlinear relationship between speed and achievable throughput. The term $\left( \frac{v_{\mathrm{p}}}{30} \right)^{\alpha}$, which is a speed factor, approximates the speed-flow relationships in the Highway Capacity Manual \cite{national1930highway}, and traffic flow theory literature \cite{daganzo1997fundamentals}. While this speed factor is a power model, the user can easily switch to other popular models, such as the Greenshields model \cite{greenshields1935study} or the Greenberg model \cite{greenberg1959analysis}. The user can also change the vehicle model to increase the variety of vehicles in traffic and adjust the gap between vehicles to mimic randomness in the physical world.

    The behavior of surrounding vehicles is governed by a set of scripts that handle collision avoidance and responses to traffic signals. We develop raycasting for collision detection, enabling each vehicle to detect obstacles via virtual radar beams. Long-range and short-range beams at the front of the vehicle support car-following and stopping behavior, and two beams at the rear corners cover blind-spot areas. The sensor manager collects readings from these beams and passes them to the vehicle manager, which updates the acceleration and deceleration commands. In areas with complex maneuvers, such as lane changes or circular movements, the short-range beam can rotate to widen the view angle while keeping the computing cost low. In the simulation, we develop a traffic light control module that cycles through phases and manages the on-off states of each signal head. This allows the scenario to reproduce movements at intersections that contain one or more signalized approaches.

\begin{figure}[!t]
        \centering
        \subfloat[]{%
            \includegraphics[width=0.8\linewidth]{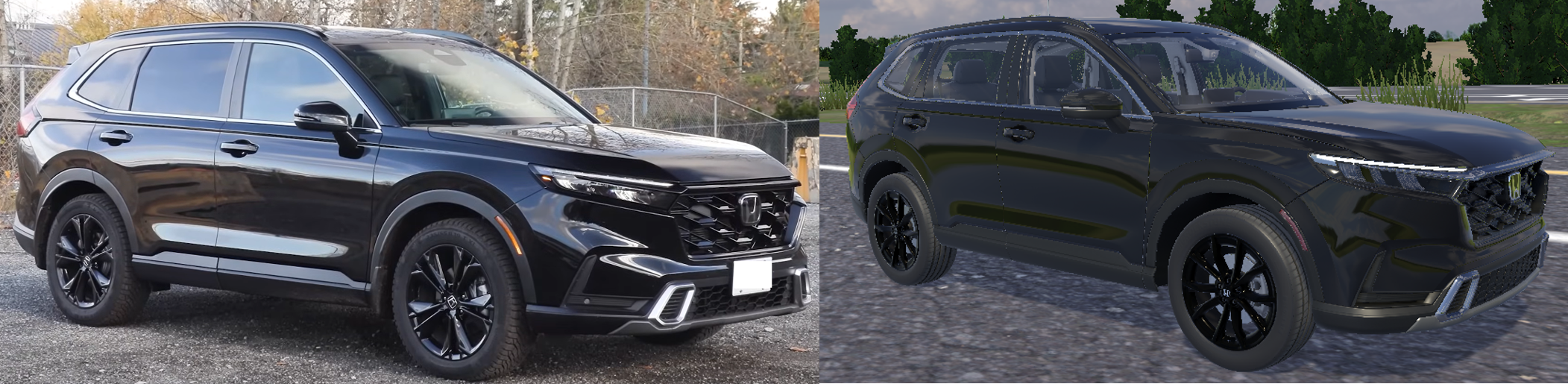}%
            \label{fig:vehicle-suv}}
        \\[0.5em]
        \subfloat[]{%
            \includegraphics[width=0.8\linewidth]{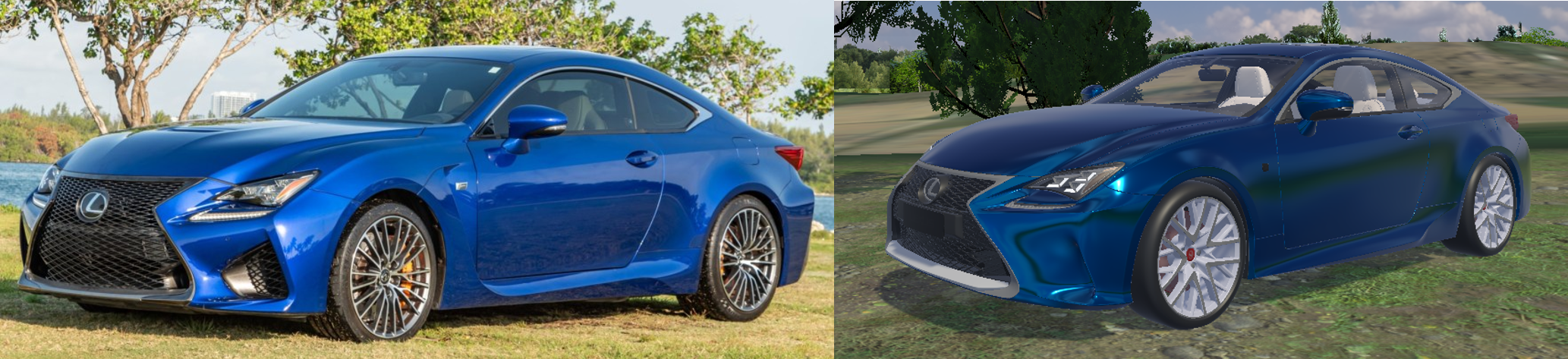}%
            \label{fig:vehicle-coupe}}
        \caption{Different types of vehicles in the real world and simulation world: (a) Honda CR-V (SUV) and (b) Lexus RC (Coupe).}
        \label{fig:vehicle-comparison}
    \end{figure}

    The ego vehicle is configured separately and follows a physics model controlled by the user. We develop the model based on a vehicle-physics framework in Unity that exposes parameters such as engine characteristics, gear ratios, suspension settings, and tire friction. These parameters can be tuned to match real vehicles, ensuring the driver experiences realistic motion when interacting with the scenario. Vehicle models are imported from external sources and adjusted in Blender and Unity to ensure proper alignment and appearance in the simulation. We simulate different types of vehicles, including but not limited to sedan, coupe, SUV, and pick-up truck. Example images of vehicles are shown in Fig.~\ref{fig:vehicle-comparison}.

\subsection{Driver Integration}
\begin{figure}[!t]
    \centering
    \includegraphics[width=1\linewidth]{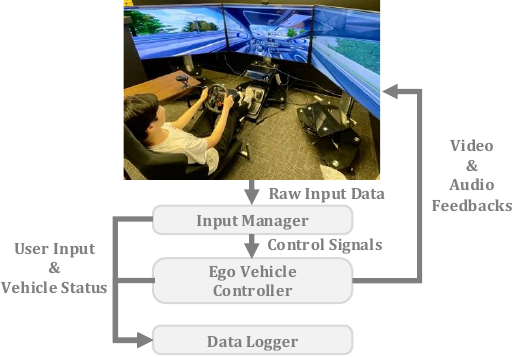}
    \caption{HITL pipeline for recording driver input and vehicle states.}
    \label{fig:Simulation Setup}
\end{figure}

    We develop the driver integration module to connect the user with the simulation world and support experiments that require direct control of the ego vehicle. The first step is the user interface, which we design to allow selection of maps, vehicles, and scenario presets. We also create guided scenarios that help users navigate innovative intersections by showing visual cues such as arrows or short text prompts. To control the ego vehicle, we integrate a steering wheel and pedal set (Logitech G29) into our system. The steering input is synchronized with the virtual steering wheel, so the driver receives a natural response when turning. For testing situations where a physical wheel is not practical, we also support keyboard and gamepad input. An example driving setup and HITL workflow are shown in Fig.~\ref{fig:Simulation Setup}.

\subsection{Performance Evaluation}
We evaluate the simulator in a standalone executable under traffic demands ranging from 0 to 2000 veh/h/l, including a 30 s warm-up period followed by 60 s of data collection. The machine we use is a Linux machine with an NVIDIA RTX 3090 Ti graphics card, an Intel i9-12900KF processor, and 64 GB of memory. As traffic demand increases, the average frame rate decreases smoothly from approximately 120 to 60 FPS, while the p95 frame time increases from approximately 9 to 23 ms. Memory usage increases by only 4.5\%, from 438 to 458 MB. CPU and GPU frame times increase from approximately 8.3 to 21.9 ms and remain within 0.03 ms of each other, indicating predictable scaling and a balanced workload between simulation and rendering.

The steps of the HITL pipeline, including device input, vehicle-physics updates, and frame rendering, are processed at an overall update frequency of 50 Hz, providing an end-to-end response latency of less than 20 ms. \textit{InterSim} can also be used to train and evaluate autonomous driving algorithms. The update frequency is configurable, and the upper limit is only capped by computational workload and hardware limits.

\section{Case Studies}
\label{sec:case study}
To demonstrate the usability of the simulation platform, we present two case studies of innovative intersections: one from the safety perspective, and another from the energy consumption perspective. All participants are informed of the experimental procedures and voluntarily agree to participate. No personally identifiable information is collected.

\subsection{Case Study I: RCI Safety}
This case study demonstrates the capability of \textit{InterSim} to support HITL safety research on innovative intersection designs. An RCI and a standard four-way intersection are constructed as comparative test environments. The experiment integrates customized intersection geometry, controlled traffic generation, human driving input, automated conflict-zone data collection, and post-processing of surrogate safety measures. The resulting comparison illustrates how the platform can be used to investigate the interaction among intersection design, traffic demand, and human driving behavior.

\subsubsection{Experiment setup} 
\begin{figure}[!t]
    \centering
    \subfloat[]{%
        \includegraphics[width=0.8\linewidth]{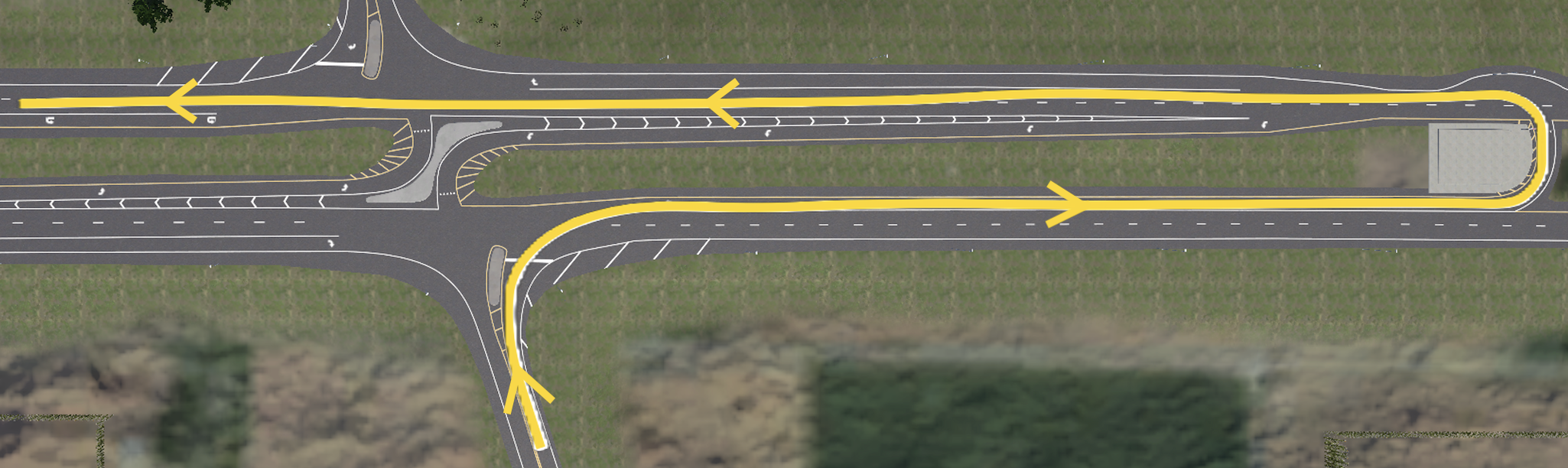}%
        \label{fig:trajectory-rci}}
    \\[0.5em]
    \subfloat[]{%
        \includegraphics[width=0.8\linewidth]{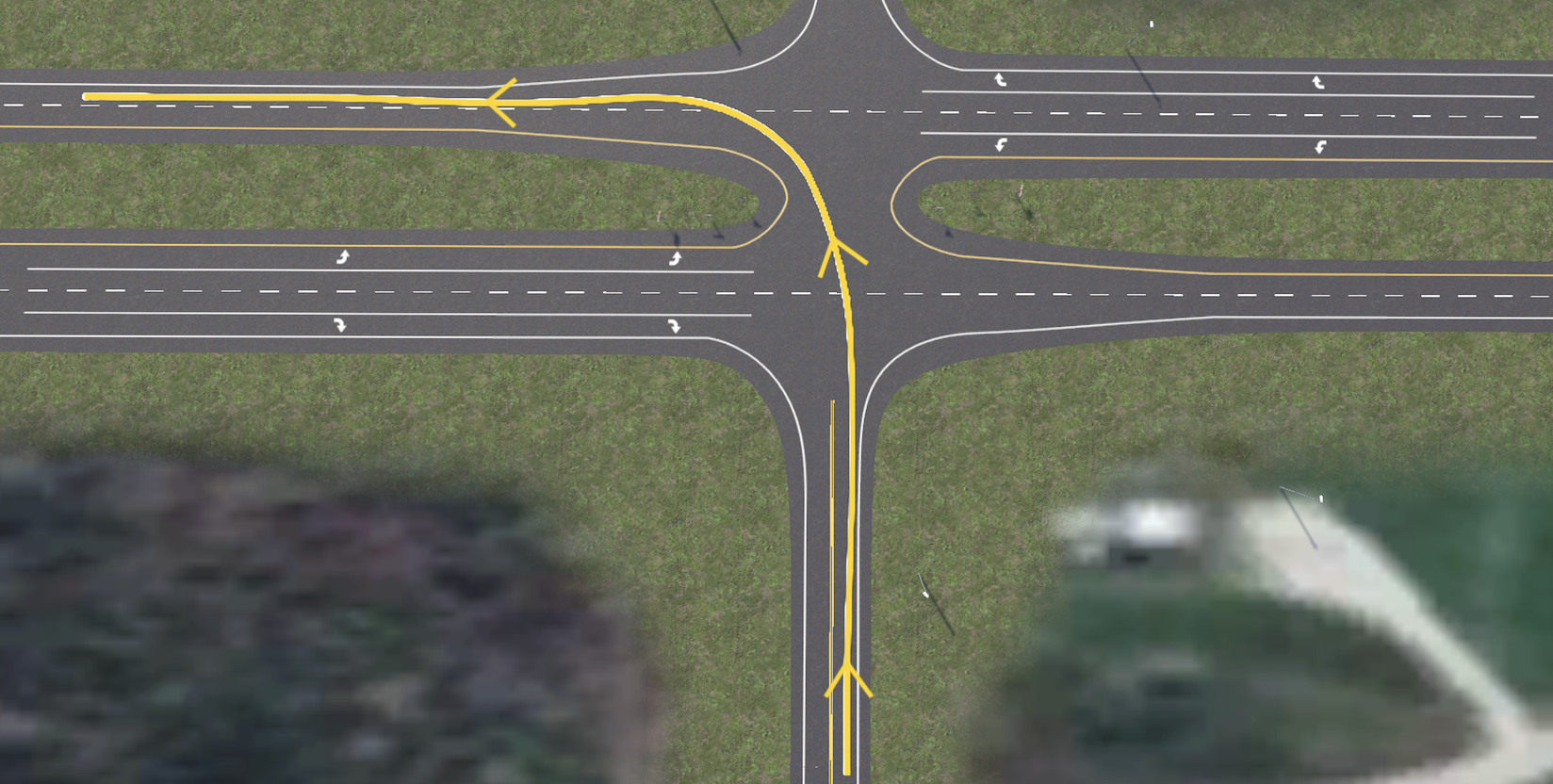}%
        \label{fig:trajectory-standard}}
    \caption{Example trajectories of experiments: (a) RCI and (b) standard intersection.}
    \label{fig:trajectories}
\end{figure}

Seven licensed drivers with one to five years of driving experience participate in the experiment, most with no prior experience navigating an RCI. Each participant completes comparable routes in both layouts, with the starting intersection randomized to reduce order and learning effects. In the RCI, participants enter the major road through a right turn and then complete a U-turn; in the standard intersection, they cross oncoming traffic through a direct left turn. All trials use the same simulated vehicle, steering wheel, and pedal setup, and example trajectories are shown in Fig.~\ref{fig:trajectories}.

Using the vehicle generation module discussed in Section~\ref{sec:workflow}, we create controlled traffic flow in both intersections, with an adjustable variation in space between vehicles, to simulate a more realistic traffic environment. The detailed setup of the traffic generation is shown in Table~\ref{tab:traffic}. Each participant drives through each map under five different traffic volumes, ranging from 200 veh/h/l to 1000 veh/h/l as shown in the table. In each map, four lanes of traffic are set to continuously generate vehicles in both directions on the major road (two lanes in each direction), with the same travel distance. Each surrounding vehicle is set up with one radar sensor facing forward to avoid rear-end collisions with the ego vehicle controlled by the user, but users might still crash into them from other directions.

\begin{figure}[!t]
    \centering
    \includegraphics[width=0.8\linewidth]{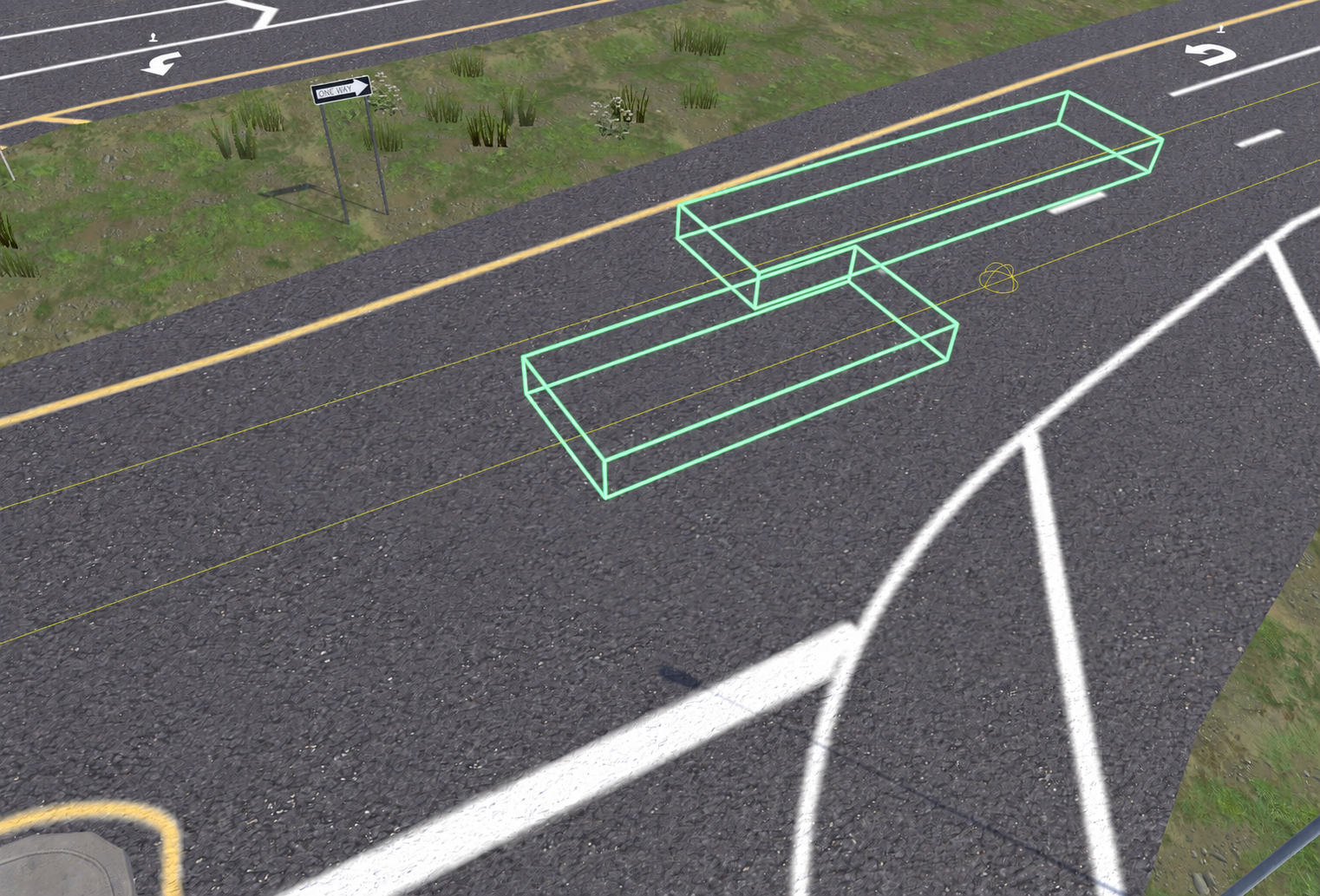}
    \caption{Example PET zones in RCI during a right turn, highlighted in green boxes.}
    \label{fig:pet}
\end{figure}

The major metric used here for safety evaluation is PET, which is commonly used in intersection studies. In this study, four conflict areas are defined for each of the intersections, where the path of a right-turn or left-turn vehicle intersects with each lane in incoming traffic on the major road, as shown in Fig.~\ref{fig:pet}. Whenever a vehicle enters or exits the PET zones, its information and timestamp are recorded. Therefore, for each PET zone, we can calculate the PET as:
    \begin{equation}
        \mathrm{PET} = t_2 - t_1
    \end{equation}
where $t_1$ is the time when the preceding vehicle is leaving the PET zone, and $t_2$ is the time when the following vehicle is entering the PET zone. To focus on critical events that might cause a collision, we choose to analyze only the PET values that are lower than five seconds, as suggested in SSAM \cite{lakshmi_pet}.

\begin{table}[!t]
    \centering
    \caption{Parameters for Traffic Generation in Case Study I}
    \begin{tabular}{ll}
    \hline
    \textbf{Parameter} & \textbf{Value} \\ 
    \hline
    Base Flow Rate $q_0$ (veh/h/l) & 200 / 400 / 600 / 800 / 1000\\
    Global Density Multiplier $\delta_{\mathrm{g}}$ & 1.0 \\
    Lane Density Multiplier $\delta_{\mathrm{l}}$ & 1.0 \\
    Speed (m/s) & 20.0 \\
    Minimum Vehicle Spacing (s) & 2.0 \\
    Spacing Variation (s) & 1.0 \\
    Lane Capacity Factor $\eta_{\mathrm{cap}}$ & 1.0 \\
    Speed Flow Exponent $\alpha$ & 1.0 \\
    \hline
    \end{tabular}
    \label{tab:traffic}
\end{table}

\subsubsection{Results and discussion} 
\begin{figure*}[t]
    \centering
    \includegraphics[width=1\linewidth]{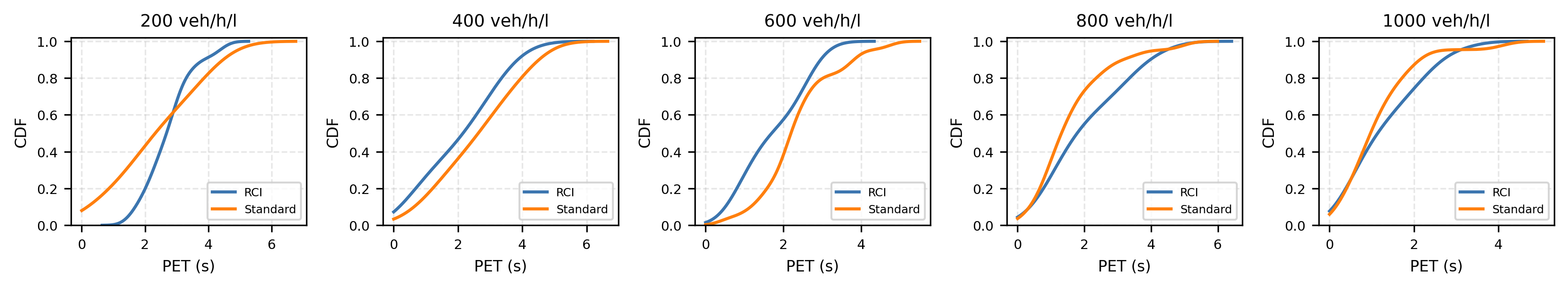}
    \caption{Gaussian-kernel-smoothed cumulative distribution curves of PET values for the RCI and standard intersections under five traffic volumes: 200, 400, 600, 800, and 1000 veh/h/l. Detailed comparisons at selected PET thresholds are provided in Table~\ref{tab:pet_ecdf_rci_vs_standard}.}
    \label{fig:pet_ecdf}
\end{figure*}

The PET data collected from the HITL driving trials are post-processed to compare their cumulative distributions across the five evaluated traffic volumes. For visual clarity, Fig.~\ref{fig:pet_ecdf} presents Gaussian-kernel-smoothed cumulative distribution curves for the recorded PET values in the RCI and standard four-way intersections. The figure highlights how the PET distributions shift across traffic demand levels, while the quantitative comparisons at selected PET thresholds are reported in Table~\ref{tab:pet_ecdf_rci_vs_standard}.

\begin{table*}[htbp]
\centering
\caption{Empirical cumulative distribution function (CDF) comparison between RCI and standard intersections at selected PET thresholds ($F(x)=P(\mathrm{PET}\leq x)$). $\Delta = F_{\mathrm{RCI}}(x)-F_{\mathrm{Standard}}(x)$.}
\label{tab:pet_ecdf_rci_vs_standard}

\setlength{\tabcolsep}{8pt} 
\begin{tabular}{c|ccc|ccc|ccc|ccc}
\hline
\multirow{2}{*}{Veh/h/l}
& \multicolumn{3}{c|}{$x=1\,\mathrm{s}$}
& \multicolumn{3}{c|}{$x=1.5\,\mathrm{s}$}
& \multicolumn{3}{c|}{$x=2\,\mathrm{s}$}
& \multicolumn{3}{c}{$x=3\,\mathrm{s}$} \\
\cline{2-13}
& RCI & Standard & $\Delta$
& RCI & Standard & $\Delta$
& RCI & Standard & $\Delta$
& RCI & Standard & $\Delta$ \\
\hline
200  & 0.000 & 0.200 & -0.200 & 0.000 & 0.200 & -0.200 & 0.182 & 0.600 & -0.418 & 0.818 & 0.600 & 0.218 \\
400  & 0.280 & 0.125 &  0.155 & 0.360 & 0.250 &  0.110 & 0.440 & 0.375 &  0.065 & 0.760 & 0.542 & 0.218 \\
600  & 0.333 & 0.091 &  0.242 & 0.500 & 0.182 &  0.318 & 0.542 & 0.364 &  0.178 & 0.958 & 0.818 & 0.140 \\
800  & 0.333 & 0.381 & -0.048 & 0.481 & 0.524 & -0.042 & 0.556 & 0.762 & -0.206 & 0.741 & 0.905 & -0.164 \\
1000 & 0.500 & 0.545 & -0.045 & 0.600 & 0.773 & -0.173 & 0.750 & 0.864 & -0.114 & 0.950 & 0.955 & -0.005 \\
\hline
\end{tabular}
\end{table*}

Four PET thresholds commonly used in previous studies are selected for comparison: $x = 1~\mathrm{s}$~\cite{lakshmi_pet, ismail2009automated}, $x = 1.5~\mathrm{s}$~\cite{fu2016pedestrian, chaudhari2021exploring, zangenehpour2016signalized}, $x = 2~\mathrm{s}$~\cite{ALHAJYASEEN201266, marisamynathan2020pedestrian}, and $x = 3~\mathrm{s}$~\cite{lord1996analysis, gettman2008surrogate}. Table~\ref{tab:pet_ecdf_rci_vs_standard} reports the cumulative probability $F(x) = P(\mathrm{PET} \leq x)$ for each intersection and traffic volume, together with the difference $\Delta = F_{\mathrm{RCI}}(x) - F_{\mathrm{Standard}}(x)$. A negative $\Delta$ indicates a lower observed proportion of PET values below the selected threshold for the RCI, whereas a positive $\Delta$ indicates a lower proportion for the standard intersection.

The results indicate that the relative safety performance of the two layouts varies with traffic demand. At 200 veh/h/l, few conflicts are observed, resulting in a limited number of PET observations at this traffic level. The RCI records no PET values below 1 s or 1.5 s, whereas the corresponding cumulative probabilities for the standard intersection are both 0.200. The largest difference occurs at $x = 2~\mathrm{s}$, where the RCI has a lower cumulative probability than the standard intersection ($\Delta = -0.418$).

In contrast, at moderate traffic volumes of 400--600 veh/h/l, this trend reverses, and the standard intersection records lower cumulative probabilities than the RCI across all evaluated thresholds. This gap peaks at 600 veh/h/l, reaching $\Delta = 0.318$ at $x = 1.5~\mathrm{s}$. The higher proportion of short PETs in the RCI stems from its two-stage maneuver: after turning right onto the major road, drivers must weave across traffic to reach the downstream U-turn, leading them to accept smaller temporal gaps when moderate headways are available.

At higher traffic volumes of 800--1000 veh/h/l, the observed PET distributions generally favor the RCI. At 800 veh/h/l, the largest difference occurs at $x = 2~\mathrm{s}$, where 76.2\% of interactions in the standard intersection fall below the threshold, compared with 55.6\% in the RCI ($\Delta = -0.206$). At 1000 veh/h/l, the RCI also records lower cumulative probabilities at the 1 s and 1.5 s thresholds, with $\Delta = -0.045$ and $\Delta = -0.173$, respectively. These findings suggest that replacing a direct multi-lane crossing maneuver with a right-turn and U-turn sequence may provide larger temporal clearances under high traffic demand.

These demand-dependent results demonstrate that \textit{InterSim} can capture variations in human driving behavior and surrogate safety outcomes across different intersection layouts and traffic conditions. Returning these HITL safety evaluations through the digital twin cycle enables IOOs to identify potential conflict risks under projected traffic demands and refine candidate RCI designs before physical deployment.

\subsection{Case Study II: Eco-driving}
To illustrate the HITL capability of the simulator, we design a second case study comparing the energy consumption between a human driver and a rule-based agent driving through a signalized intersection. We implement a trigonometry-based eco-driving agent in our \textit{InterSim} simulator, letting it drive through an intersection with fewer speed changes based on SPaT received from V2I communications, and compare its energy consumption with a human driver driving in the same scenario. The results demonstrate \textit{InterSim}'s ability to support controlled comparisons between HITL driving and V2I-enabled automated control, while enabling energy-consumption analysis at both the route and trajectory levels.

\subsubsection{Rule-based agent} The rule-based agent adapts the eco-approach and departure (EAD) controller proposed by Wang et al.~\cite{wang2019cooperativeeco} for use as a single-agent baseline in \textit{InterSim}. The agent receives SPaT information and the distance to the stop line through \textit{InterSim}'s V2I module and uses them with its current speed to select a cruise, smooth speed-adjustment, or stop-and-go trajectory.

The nominal cruising time-to-arrival is calculated as
\begin{equation}
    t_{\mathrm{cr}} = \frac{d_s}{v_s},
\end{equation}
where $d_s$ and $v_s$ denote the current distance to the stop line and vehicle speed, respectively. The current traffic-signal phase obtained from the SPaT message is denoted by $\sigma \in \{\mathrm{green},\mathrm{yellow},\mathrm{red}\}$. Based on $\sigma$ and the remaining phase durations received, the available green interval is represented as
\begin{equation}
\mathcal{G} =
\begin{cases}
[0,g_e^{\mathrm{cur}}] \cup
[g_s^{\mathrm{next}},g_e^{\mathrm{next}}],
& \sigma=\mathrm{green}, \\[2mm]
[g_s^{\mathrm{next}},g_e^{\mathrm{next}}],
& \sigma\in\{\mathrm{yellow},\mathrm{red}\},
\end{cases}
\end{equation}
where $g_e^{\mathrm{cur}}$ is the remaining duration of the current green phase, and $g_s^{\mathrm{next}}$ and $g_e^{\mathrm{next}}$ denote the start and end times of the next green interval relative to the current time. The agent also estimates the earliest and latest feasible arrival times, $t_e$ and $t_l$, from the speed, acceleration, coasting-speed, and jerk constraints. These values are compared with $\mathcal{G}$ to determine the trajectory mode, as summarized in Algorithm~\ref{alg:eco_agent}.

\begin{algorithm}[t]
\caption{Single-Agent Eco-Driving Decision Logic}
\label{alg:eco_agent}
\small
\begin{algorithmic}[1]

\Require Distance $d_s$, speed $v_s$, signal phase $\sigma$,
         SPaT information, and vehicle constraints
\Ensure Trajectory mode, arrival time $t_a$, and reference acceleration $a_{\mathrm{ref}}$

\State Compute nominal arrival time $t_{\mathrm{cr}}=d_s/v_s$
\State Determine available green interval $\mathcal{G}$
\State Estimate feasible arrival bounds $(t_e,t_l)$

\If{$d_s \leq 0$}
    \State Resume free-flow acceleration
\ElsIf{$t_{\mathrm{cr}}\in\mathcal{G}$}
    \State Select cruise mode
    \State $t_a\gets t_{\mathrm{cr}},\quad a_{\mathrm{ref}}\gets0$
\ElsIf{$[t_e,t_l]\cap\mathcal{G}\neq\emptyset$}
    \State Select a feasible $t_a$ within the green interval
    \State Generate a smooth trigonometric speed-adjustment profile
\Else
    \State Set $t_a$ to the beginning of the next green interval
    \State Generate a stop-and-go trajectory
\EndIf

\end{algorithmic}
\end{algorithm}

\subsubsection{Experiment setup} 
\begin{figure}[!t]
    \centering
    \includegraphics[width=1\linewidth]{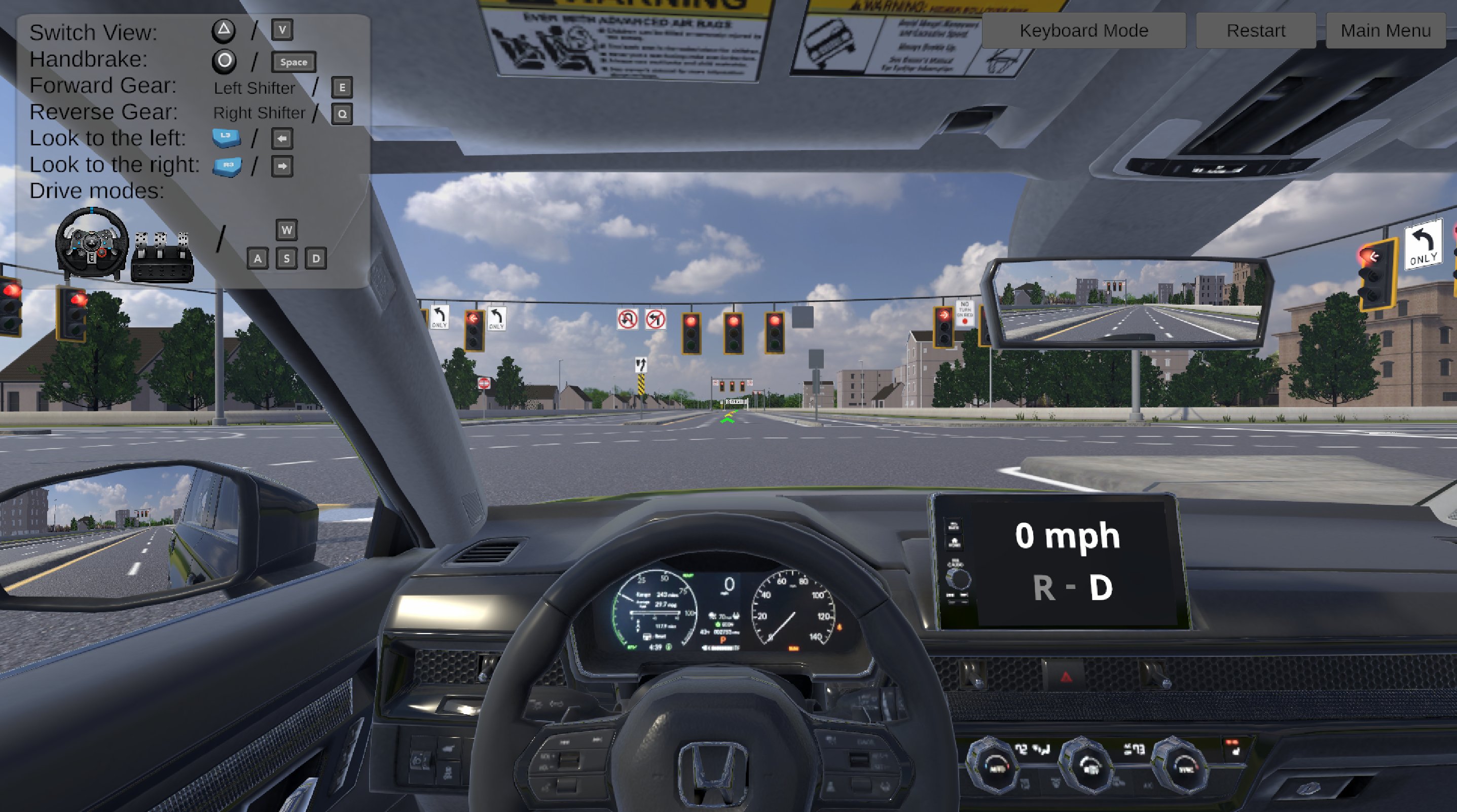}
    \caption{In-cabin view of a driver waiting at a traffic light in case study II.}
    \label{fig:ecodrive}
\end{figure}

After implementing the rule-based agent algorithm in C\#, we set up the driving scenario in the DLT map in \textit{InterSim}, as it has a long straight road crossing the signalized intersection, allowing sufficient space to evaluate the eco-driving behavior. Both human and agent drivers start at the same place before the traffic light, and finish at the same place after the traffic light. Seven participants with different driving styles are involved in the experiment. Each human driver runs 10 trips across the intersection, and the agent runs 70 trips. For each driver, the traffic light starts from the same initial state and follows the same phase cycle until the end of all 10 trips. In this case, the drivers experience different states when they arrive at the traffic light. No other traffic is generated in this case study to minimize the impact of surrounding traffic on both human and rule-based agent drivers. Fig.~\ref{fig:ecodrive} demonstrates the in-cabin view of a human driver waiting at the traffic light of the DLT.

To estimate energy consumption and pollutant emissions, we use MOVESTAR, an open-source model developed from the U.S. EPA MOVES framework \cite{wang2020movestar,USEPA_MOVES5_2024}. We choose ``passenger vehicle'' as the vehicle type, and use MOVESTAR to calculate Vehicle Specific Power (VSP) based on the recorded vehicle speed and acceleration, determine the corresponding operating mode, and estimate energy and emission rates at each time step. The key metrics chosen from MOVESTAR are energy consumption, CO, CO$_2$, HC, and NO$_x$ emissions for each trip.

\subsubsection{Results and discussion}
As the eco-driving algorithm is designed to improve energy consumption before the traffic light, 70 human driver trips and 70 rule-based agent trips are compared over the 300 m approaching section.

Table~\ref{tab:approach_energy_emissions} summarizes the normalized mean value of metrics over all trips. The agent records a mean energy intensity of 2308.34 kJ/km, compared with 3182.79 kJ/km for the human trips, corresponding to an estimated difference of $-27.5\%$. The average difference for the pollutant estimates ranges from $-27.5\%$ for CO$_2$ to $-52.5\%$ for NO$_x$.

\begin{table}[!t]
    \centering
    \caption{Mean per-trip energy and emission metrics within the 300 m approaching section, normalized by distance. The estimated difference is
    $(\mathrm{Agent}-\mathrm{Human})/\mathrm{Human}\times100\%$.}
    \label{tab:approach_energy_emissions}
    \renewcommand{\arraystretch}{1.15}
    \begin{tabular*}{\columnwidth}{
        @{\extracolsep{\fill}} l r r r @{}
    }
        \toprule
        Metric &
        \shortstack{Human\\mean} &
        \shortstack{Agent\\mean} &
        \shortstack{Estimated \\
        difference (\%)} \\
        \midrule
        Energy (kJ/km) & 3182.79 & 2308.34 & $-27.5$ \\
        CO (g/km)      & 4.7915  & 2.8401  & $-40.7$ \\
        CO$_2$ (g/km)  & 228.75  & 165.90  & $-27.5$ \\
        HC (g/km)      & 0.20868 & 0.14682 & $-29.6$ \\
        NO$_x$ (g/km)  & 0.58726 & 0.27919 & $-52.5$ \\
        \bottomrule
    \end{tabular*}
\end{table}

The earlier and more gradual speed adjustment of the agent is consistent with its use of SPaT information when planning the approach trajectory. Beyond individual vehicle energy use, eco-driving helps smooth vehicle arrivals and reduce unnecessary braking at signalized intersections. With surrounding traffic and different penetration rates, \textit{InterSim} can further examine whether these trajectory changes contribute to more stable traffic flow and shorter queues. For IOOs, these findings can inform decisions on deploying roadside V2I infrastructure or adjusting signal timing plans to reduce emissions at intersections, especially with the help of human feedback.

\section{Conclusion} 
\label{sec:conclusion} 
In this paper, we developed \textit{InterSim}, a Unity game engine-based HITL driving simulator that constructed digital twins of innovative intersections and evaluated them with human and automated drivers. We built digital twins of five intersection types from the documents and data of IOOs. The digital twin cycle returned safety and emission analyses to those operators to guide pre-deployment design refinements, while the HITL cycle placed a human driver or an automated driving algorithm in the ego vehicle in real time. The platform integrated configurable traffic, sensors, driver interfaces, and data collection, while maintaining real-time operation under different traffic demands. Two case studies further demonstrated the research capabilities of \textit{InterSim}. The RCI study showed that the platform supported HITL surrogate-safety analysis across different traffic demands, providing IOOs with pre-deployment evidence to identify conflict risks and refine candidate layouts. The eco-driving study demonstrated its ability to compare human drivers with a V2I-enabled automated agent, offering insights to help IOOs evaluate roadside V2I deployment and signal timing adjustments for emission reduction. Together, these studies showed that \textit{InterSim} provided a practical testbed for both researchers and IOOs to evaluate and improve innovative intersection designs before field implementation. Future work will focus on automating the map-generation process, expanding the range of supported scenarios and human-subject studies, and closing the digital twin cycle in real time by updating the digital twins with live field data.

\section*{Acknowledgment}
This work was supported by the Indiana Department of Transportation (INDOT) Joint Transportation Research Program (JTRP) under Grant JTRP-4932. Special thanks to INDOT members including Dan McCoy, Taylor Ruble, Shuo Li, David Reamer, Cat Seely, Jennifer Beck, and Tom Ford, for their precious advice. Special thanks to Purdue Envision Center lab members including George Takahashi, Angel Rivera Cardenas, Blake Edwards, and David Chen, for their technical support.

\bibliographystyle{IEEEtran}
\bibliography{main}
\begin{IEEEbiography}[{\includegraphics[width=1in,height=1.25in,clip,keepaspectratio]{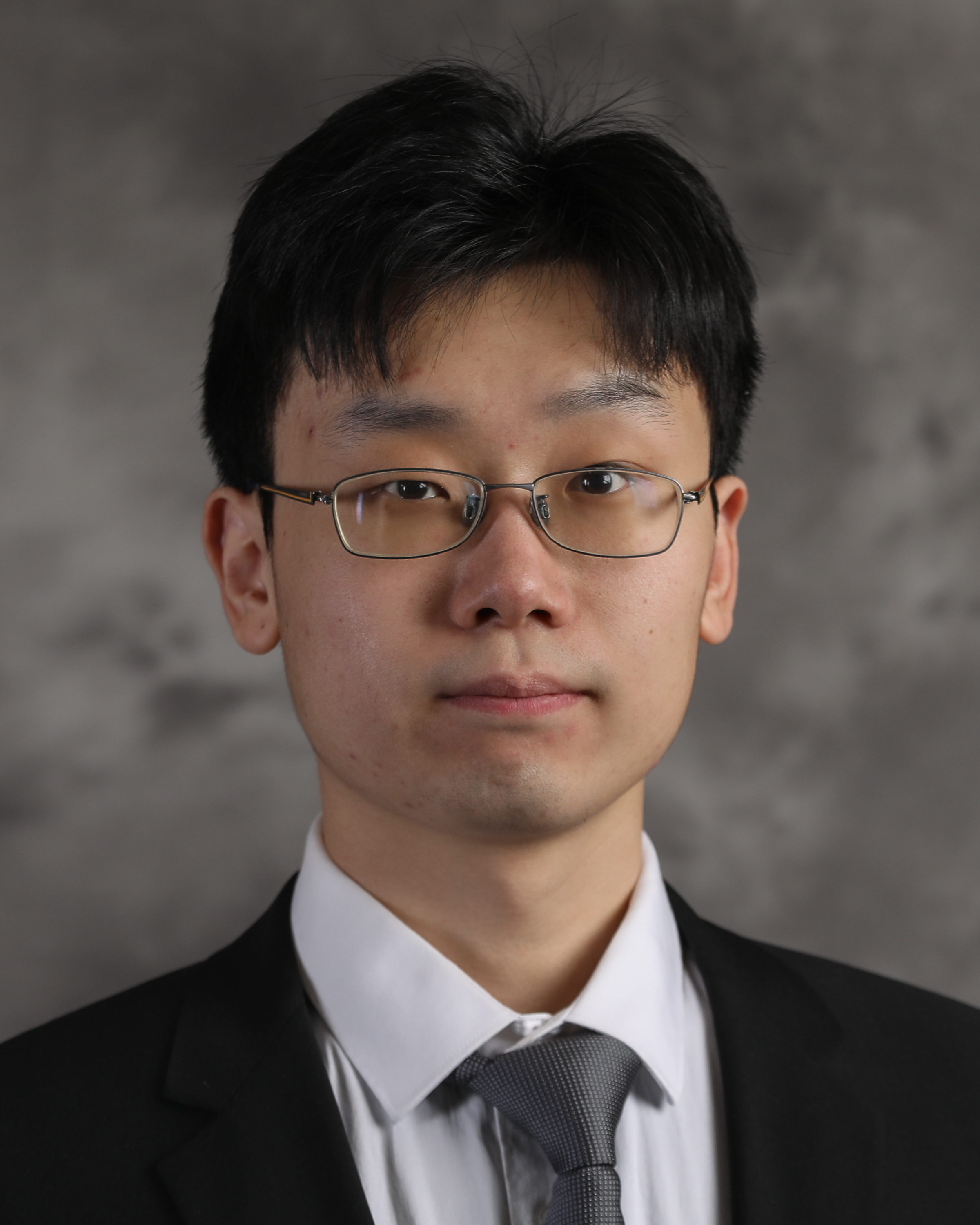}}]{Luyang Jiang}
received the B.Eng. degree in computer engineering from the University of British Columbia, Vancouver, BC, Canada. He is currently pursuing the M.S. degree in electrical and computer engineering with the Purdue Digital Twin Lab, Purdue University, West Lafayette, IN, USA. His research interests include autonomous driving, computer vision, machine learning, and driving simulation.
\end{IEEEbiography}

\begin{IEEEbiography}[{\includegraphics[width=1in,height=1.25in,clip,keepaspectratio]{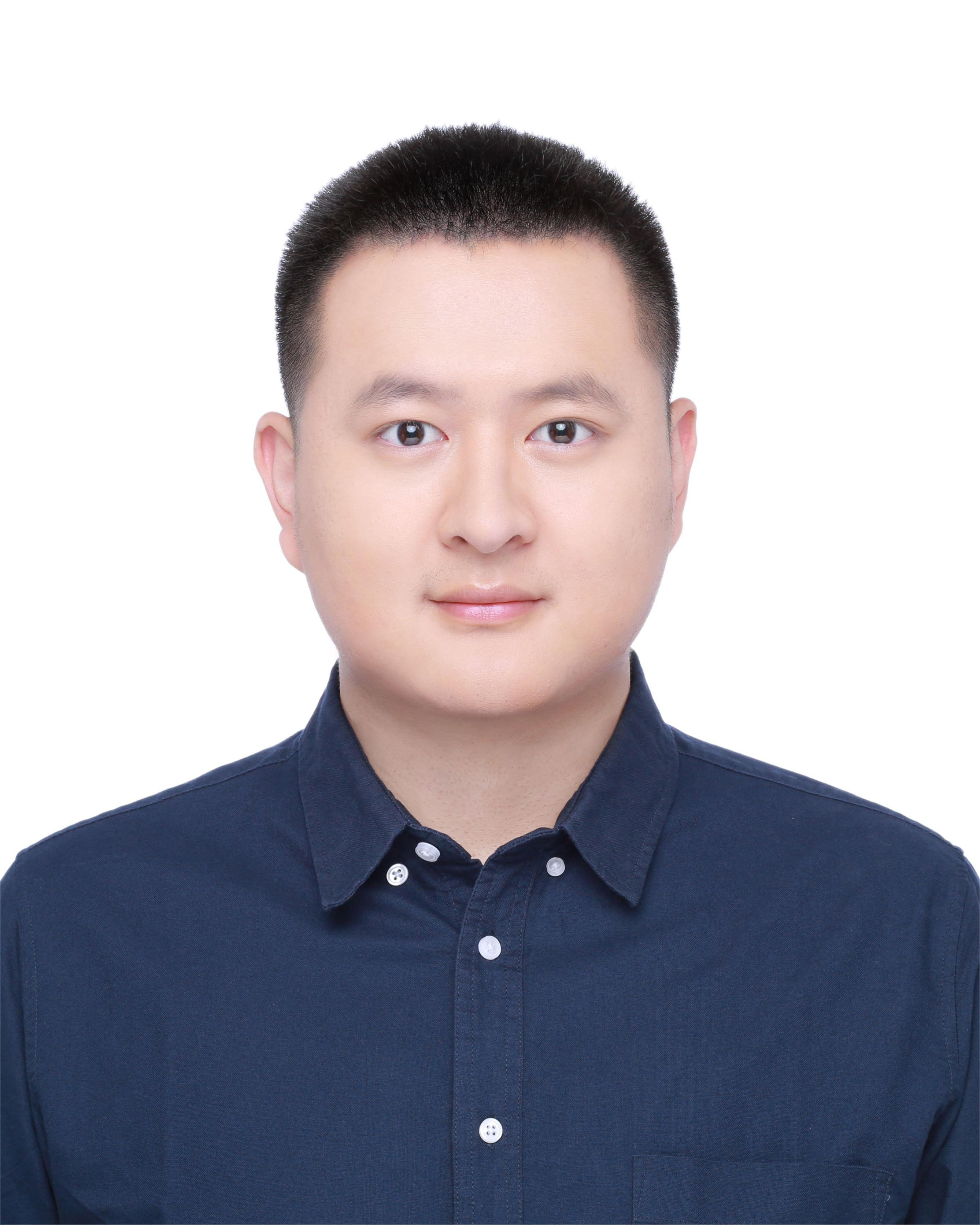}}]{Chuheng Wei}
received the B.S. degree from the Communication University of China, Beijing, China, in 2017, the M.S. degree in smart, connected and autonomous vehicles from the University of Warwick, Coventry, U.K., in 2022, and the Ph.D. degree in electrical and computer engineering from the University of California at Riverside, Riverside, CA, USA, in 2026. He is a Postdoctoral Researcher with the College of Engineering, Purdue University, West Lafayette, IN, USA. His research interests include multimodal perception in autonomous driving, V2X cooperative perception, and foundation models.
\end{IEEEbiography}

\begin{IEEEbiography}[{\includegraphics[width=1in,height=1.25in,clip,keepaspectratio]{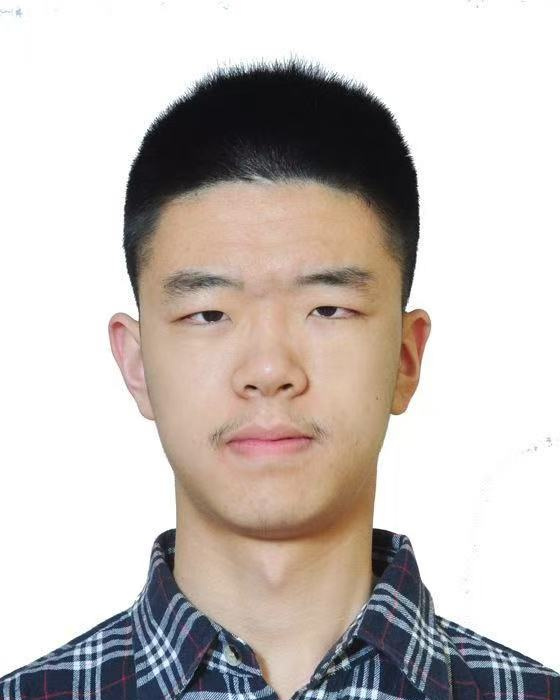}}]{Yaguang Li}
received the B.S. degree in automation from Zhejiang University, Hangzhou, China, in 2023, and the M.S. degree in electrical and computer engineering from Carnegie Mellon University, Pittsburgh, PA, USA. He is currently pursuing the Ph.D. degree in electrical and computer engineering with the Purdue Digital Twin Lab, Purdue University, West Lafayette, IN, USA. His research interests include control and learning theory, autonomous driving, and robotics.
\end{IEEEbiography}

\begin{IEEEbiography}[{\includegraphics[width=1in,height=1.25in,clip,keepaspectratio]{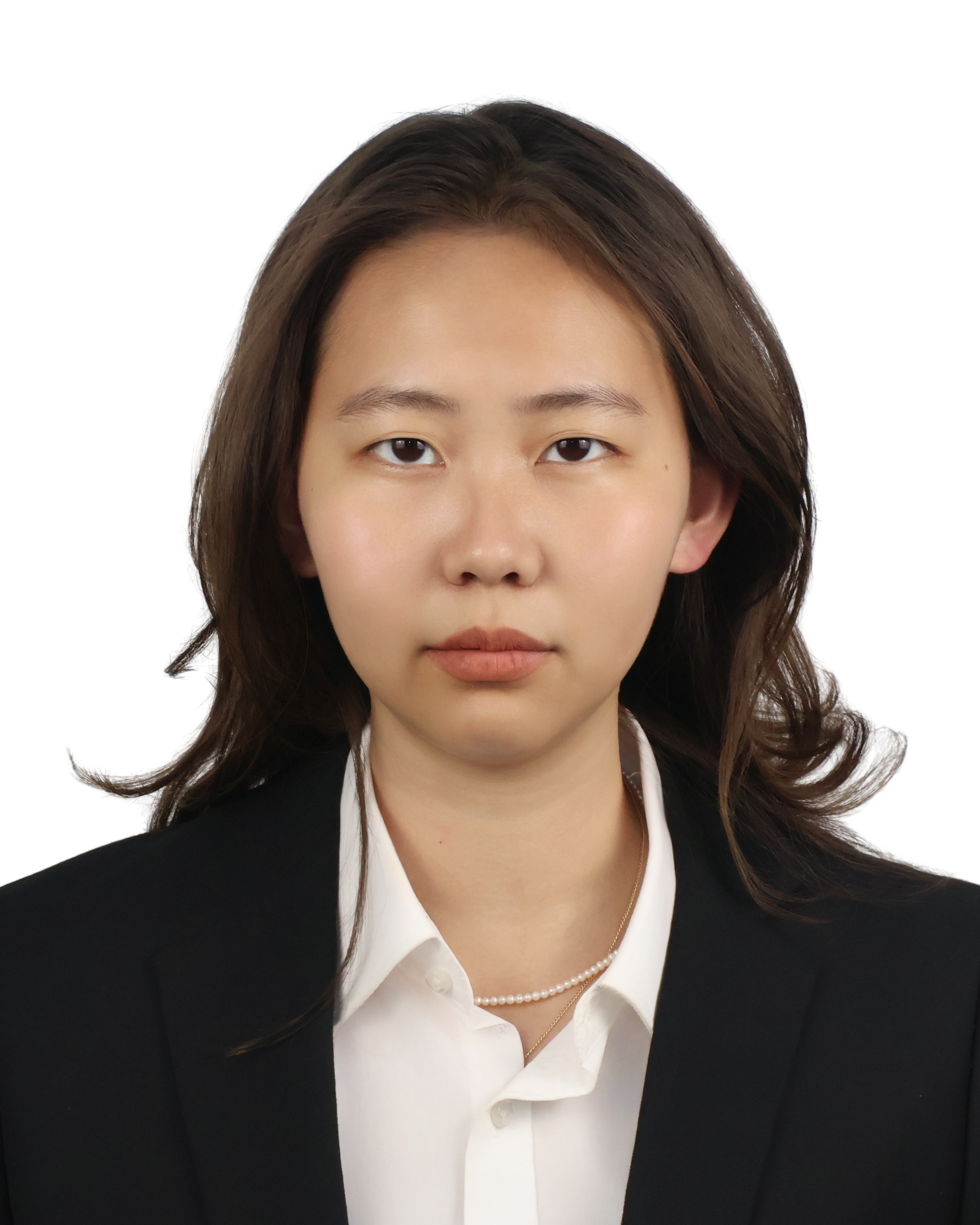}}]{Shaoqi Wang}
received the B.S. degree in vehicle engineering from Tongji University, Shanghai, China, in 2026. She is currently with the College of Engineering, Purdue University, West Lafayette, IN, USA. Her research interests include intelligent transportation systems, driving simulation, and autonomous driving.
\end{IEEEbiography}

\begin{IEEEbiography}[{\includegraphics[width=1in,height=1.25in,clip,keepaspectratio]{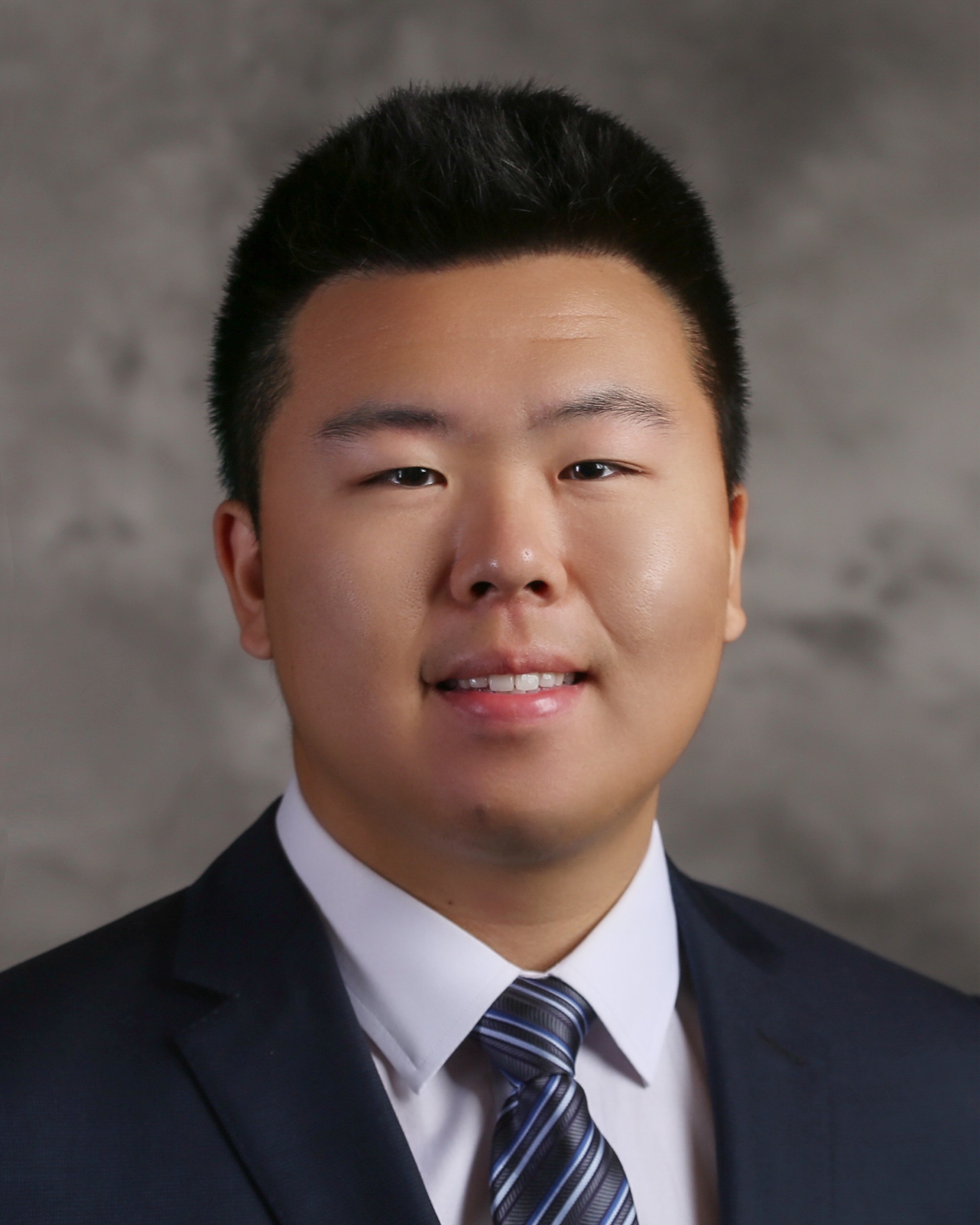}}]{Yupeng Zhou}
received the B.S. degree in mechanical engineering from the South China University of Technology, Guangzhou, China, in 2022, and the M.S. degree in mechanical engineering from Purdue University, West Lafayette, IN, USA, where he is currently pursuing the Ph.D. degree in mechanical engineering. His research interests include robotics, control systems, and autonomous vehicles.
\end{IEEEbiography}

\begin{IEEEbiography}[{\includegraphics[width=1in,height=1.25in,clip,keepaspectratio]{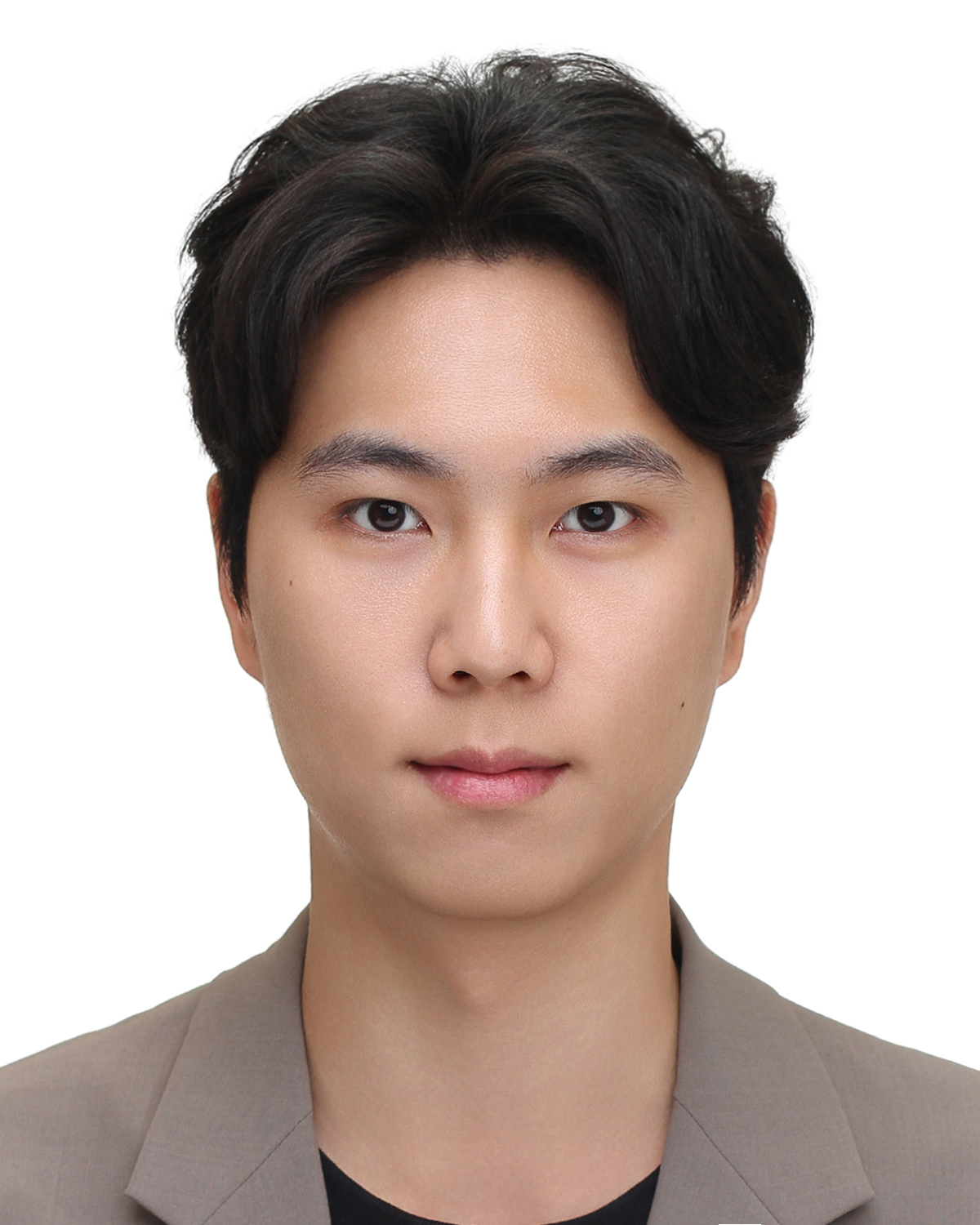}}]{Hyunsoo Yun}
received the B.S. degree in civil engineering from Kyung Hee University, Seoul, South Korea, in 2020, and the Ph.D. degree in civil and environmental engineering from Seoul National University, Seoul, in 2025. He is currently a Postdoctoral Researcher with the Lyles School of Civil and Construction Engineering, Purdue University, West Lafayette, IN, USA. His research interests include travel demand modeling, agent-based simulation, shared mobility systems, and reinforcement learning in transportation.
\end{IEEEbiography}

\begin{IEEEbiography}[{\includegraphics[width=1in,height=1.25in,clip,keepaspectratio]{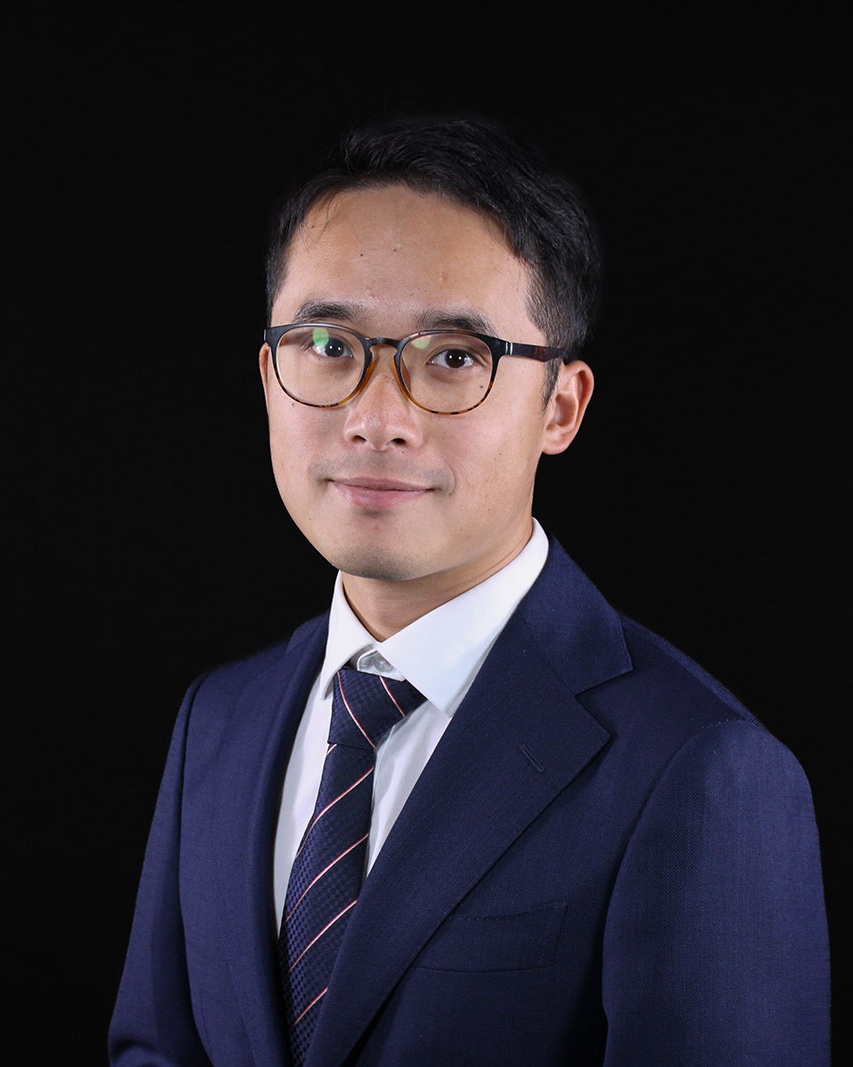}}]{Xishun Liao}
(Member, IEEE) received the B.Eng. degree from the Beijing University of Posts and Telecommunications, Beijing, China, in 2016, the M.Eng. degree in mechanical engineering from the University of Maryland, College Park, MD, USA, in 2018, and the Ph.D. degree in electrical and computer engineering from the University of California at Riverside, Riverside, CA, USA, in 2023. Following his Ph.D., he was a Postdoctoral Scholar with the University of California at Los Angeles (UCLA), Los Angeles, CA, USA. He is currently an Assistant Professor with the Department of Civil, Environmental, and Construction Engineering, University of Central Florida, Orlando, FL, USA. His research interests include human-centered intelligent mobility systems, transportation digital twins, and connected and automated vehicles.
\end{IEEEbiography}

\begin{IEEEbiography}[{\includegraphics[width=1in,height=1.25in,clip,keepaspectratio]{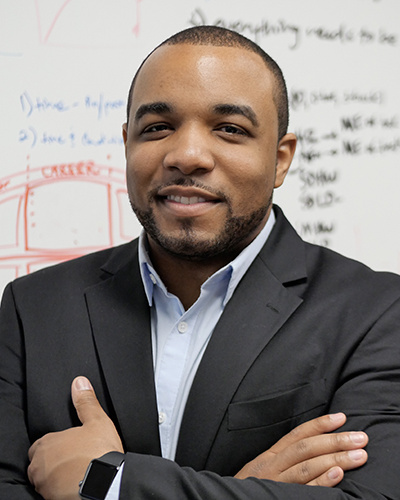}}]{Brandon Pitts}
received the B.S. degree in industrial engineering from Louisiana State University, Baton Rouge, LA, USA, in 2010, and the M.S.E. and Ph.D. degrees in industrial and operations engineering from the University of Michigan, Ann Arbor, MI, USA, in 2013 and 2016, respectively. He is currently an Associate Professor with the Edwardson School of Industrial Engineering, Purdue University, West Lafayette, IN, USA, where he directs the Next-generation Human-systems and Cognitive Engineering (NHanCE) Lab. His research interests include human factors, cognitive engineering, human-automation interaction, and autonomous driving.
\end{IEEEbiography}

\begin{IEEEbiography}[{\includegraphics[width=1in,height=1.25in,clip,keepaspectratio]{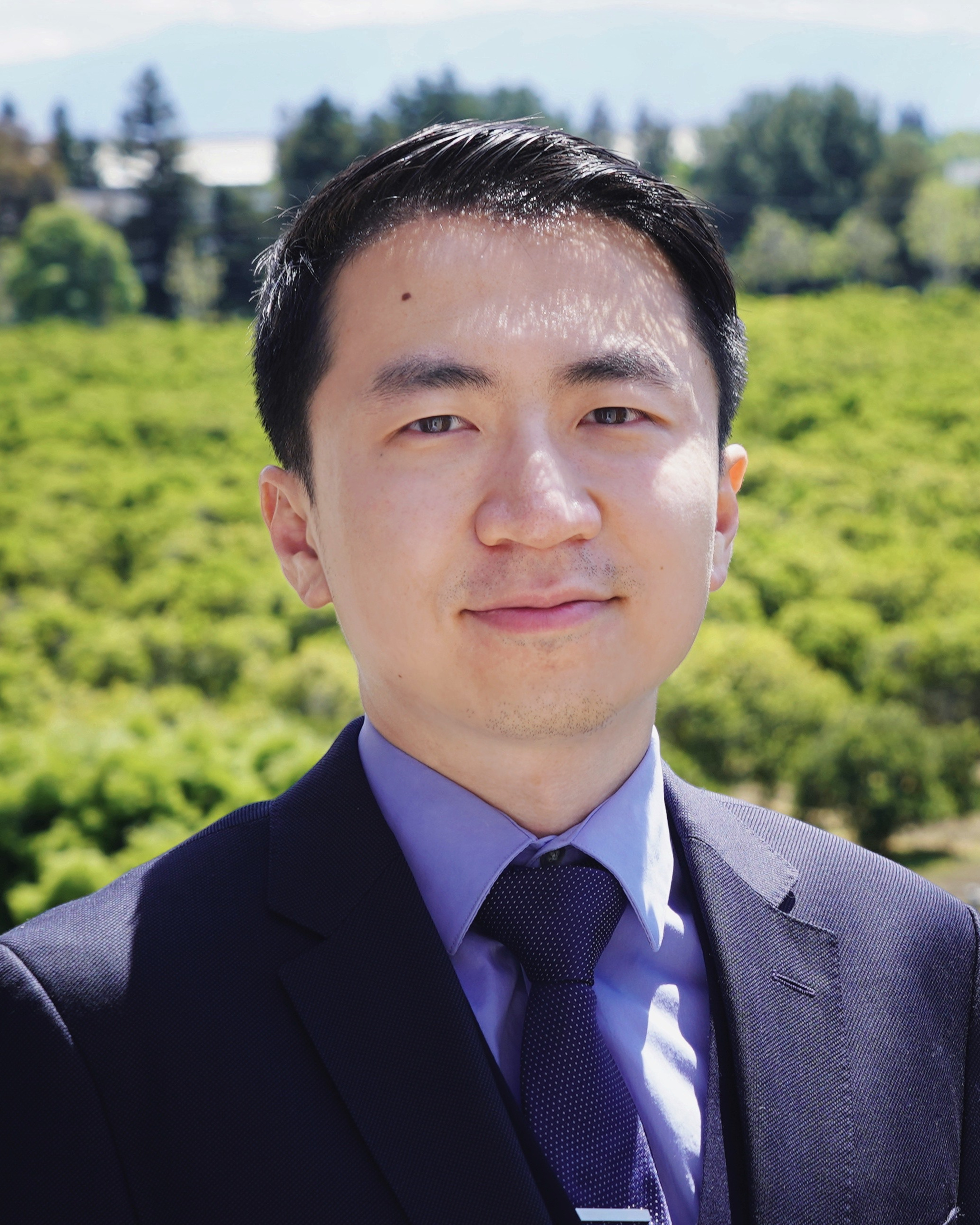}}]{Ziran Wang}
(Senior Member, IEEE) received the B.E. degree from the Beijing University of Posts and Telecommunications, Beijing, China, in 2015, and the Ph.D. degree in mechanical engineering from the University of California at Riverside, Riverside, CA, USA, in 2019. Prior to joining Purdue University in 2023, he was a Principal Researcher with Toyota Motor North America R\&D, Mountain View, CA, USA. He is currently an Assistant Professor with the Lyles School of Civil and Construction Engineering and the Elmore Family School of Electrical and Computer Engineering (by courtesy), Purdue University, West Lafayette, IN, USA, where he directs the Purdue Digital Twin Lab. His research interests include autonomous driving, digital twins, human-autonomy teaming, and intelligent transportation systems.
\end{IEEEbiography}

\end{document}